\newcommand{\ud}[2]{\mbox{$^{+ #1}_{- #2}$}}

\documentclass[preprint]{aastex}

\slugcomment{ Accepted by AJ, to appear in April 2001 }

\shortauthors{Brandt et al.}
\shorttitle{GHRS Observations of AB Dor }

\begin{document}
\title{AB Dor in '94: I. HST/GHRS Observations of the Quiescent Chromosphere of
an Active Star}

\author{J.C. Brandt\altaffilmark{1,2}}
\affil{Laboratory for Atmospheric and Space Physics,
Campus Box 392, University of Colorado, Boulder, CO 80309-0392}

\email{jbrandt@lodestar.phys.unm.edu}

\author{S.R. Heap\altaffilmark{1}}
\affil{Laboratory for Astronomy and Solar Physics,
Code 681 NASA/Goddard Space Flight Center, Greenbelt, MD 20771}
\email{hrsheap@stars.gsfc.nasa.gov}

\author{F.M. Walter\altaffilmark{1}}
\affil{State University of New York, Physics and Astronomy,
Stony Brook, NY 11794-3800}
\email{fwalter@astro.sunysb.edu}

\author{E.A. Beaver\altaffilmark{1}}
\affil{Center for Astrophysics and Space Sciences C-0111,
University of California, San Diego, La Jolla, CA 92093-0111}
\email{ebeaver@ucsd.edu}

\author{A. Boggess\altaffilmark{1}}
\affil{2420 Balsam Dr. Boulder CO 80304}
\email{boggess@lyrae.colorado.edu}

\author{K.G. Carpenter\altaffilmark{1}}
\affil{Laboratory for Astronomy and Solar Physics,
Code 681 NASA/Goddard Space Flight Center, Greenbelt, MD 20771}
\email{carpenter@stars.gsfc.nasa.gov}

\author{D.C. Ebbets\altaffilmark{1}}
\affil{Ball Aerospace \& Technologies Corp.,
P. O. Box 1062, AR1, Boulder, CO 80306}
\email{debbets@ball.com}

\author{J.B. Hutchings\altaffilmark{1}}
\affil{Dominion Astrophysical Observatory,
5071 West Saanich Road, Victoria, BC, Canada V8X~4M6}
\email{john.hutchings@hia.nrc.ca}

\author{M. Jura\altaffilmark{1}}
\affil{Department of Physics and Astronomy, University of California,
Los Angeles, CA 90095-1562}
\email{jura@clotho.astro.ucla.edu}

\author{D.S. Leckrone\altaffilmark{1}}
\affil{Laboratory for Astronomy and Solar Physics,
Code 681 NASA/Goddard Space Flight Center, Greenbelt, MD 20771}
\email{hrsleckrone@hrs.gsfc.nasa.gov}

\author{J.L. Linsky\altaffilmark{1}}
\affil{JILA, University of Colorado and National Institute of Standards and
Technology, Boulder, CO 80309-0440}
\email{jlinsky@jila.colorado.edu}

\author{S.P. Maran\altaffilmark{1}}
\affil{Space Sciences Directorate,
Code 600, NASA/Goddard Space Flight Center, Greenbelt MD 20771}
\email{hrsmaran@eclaire.gsfc.nasa.gov}

\author{B.D. Savage\altaffilmark{1}}
\affil{Department of Astronomy, University of Wisconsin,
475 North Charter Street, Madison, WI 53706}
\email{bdsavage@facstaff.wisc.edu}

\author{A.M. Smith\altaffilmark{1}}
\affil{Laboratory for Astronomy and Solar Physics,
Code 681 NASA/Goddard Space Flight Center, Greenbelt, MD 20771}
\email{hrssmith@hrs.gsfc.nasa.gov}

\author{L.M. Trafton\altaffilmark{1}}
\affil{MacDonald Observatory and Astronomy Department,
University of Texas, Austin, TX 78712}
\email{lmt@astro.as.utexas.edu}

\author{R.J. Weymann\altaffilmark{1}}
\affil{Observatories of the Carnegie Institution of Washington,
813 Santa Barbara Street, Pasadena CA 91101}
\email{rjw@ociw.edu}

\author{D. Norman}
\affil{State University of New York, Physics and Astronomy,
Stony Brook, NY 11794-3800}
\email{dnorman@sbast3.ess.sunysb.edu}

\author{S. Redfield}
\affil{JILA, University of Colorado, Boulder, CO 80309-0440}
\email{sredfiel@casa.colorado.edu}

\altaffiltext{1}{GHRS Investigation Definition Team}
\altaffiltext{2}{Now at Department of Physics and Astronomy, Institute
for Astrophysics, University of New Mexico, Albuquerque, NM 87131}

\begin{abstract}
We analyze HST/GHRS spectra of AB Doradus, the prototypical ultra-rapidly
rotating K dwarf.
We observed chromospheric (\ion{Mg}{2}) and transition region (\ion{C}{2},
\ion{Si}{4}, \ion{C}{4}, and \ion{N}{5}) lines periodically throughout the
stellar rotation period, and provide a low dispersion stellar atlas of
78 emission lines.
The quiescent line profiles
of the chromospheric and transition region lines show narrow cores superposed
on very broad wings. 
The broad wings of
the \ion{Mg}{2} $k$ \& $h$ lines and of the transition region lines
can be explained by emission from gas co-rotating with the star and extending
out to near the Keplerian co-rotation radius (2.8 stellar radii).
While this is not a unique solution, it is consistent with previous studies
of H$\alpha$ emission that are naturally explained by large co-rotating
prominences. We find
no evidence for rotational modulation of the emission line fluxes.
The density diagnostics suggest that the transition region is formed
at constant pressure, with an electron density
2--3~$\times$~10$^{12}$~cm$^{-3}$ at a temperature of 3$\times$10$^4$K.
The electron pressure is about 100 times larger than that for the quiet
Sun.
The emission measure distribution shows a minimum between log(T) = 5 and 5.5. 
The \ion{Mg}{2} line exhibits three interstellar absorption components along
the 15~pc line of sight. We identify the lowest velocity component with the G
cloud, but the other components are not identified with any
interstellar clouds previously detected from other lines of sight.

\end{abstract}


\section{Introduction}

We study magnetically-active solar-like stars because
the characteristics of such stars may provide insights into the nature
of solar-like magnetic activity.
While we accept the solar paradigm, that stellar activity is a consequence of
magnetic fields,
it is not at all clear that any simple scaling of solar coronal structures
can account for the level of activity seen in the most active stars. 
Indeed, the very assumption of solar-like structures may bias our
interpretation of the observations.
Walter \& Byrne (1997), and Walter (1999)
proposed a non-solar paradigm for active stars,
based in part on observations of AB~Doradus. 
In this picture, a quasi-dipolar global magnetic field may dominate the
large-scale activity, while the hotter coronal gas is confined by
solar-like magnetic structures to a small scale height. An active stellar
atmosphere would then consist of a compact, solar-like
chromosphere/transition region/corona
and an extended, co-rotating envelope of large volume.
Ayres et al.\/ (1998) posit a similar picture for the magnetospheres of
active Hertzsprung gap giants.

Ultraviolet spectra sample the chromosphere and transition region, at
temperatures from about 10$^4$K through 2$\times$10$^{5}$K. While 
emission in the cooler lines may result in part from heating
by acoustic fluxes, the transition region line emission is
almost certainly from hot plasma trapped in magnetic loops.
Time-resolved spectra obtained over a stellar rotation
can yield important constraints on magnetic filling factors, global
asymmetries, and atmospheric scale heights. Velocity-resolved line profiles
can be inverted to yield spatial resolution on these stars, and images of their
surfaces.
Our goal, through detailed observations of highly active stars,
is to test our understanding
of stellar coronae and chromospheres, of the morphology of the magnetic
field, and of its relation to the observed activity. 

\subsection{AB Doradus}

AB~Doradus (HD~36705; K0-2 IV--V),
the brightest (V=6.7) of the ultra-rapid rotators
(P$_{rot}$=0.51479~d; $v$~sin~$i$=91 km~s$^{-1}$),
is the quintessential active young single star.
In addition to large flares (e.g., Robinson \& Cameron 1986), large
starspots (Anders et al 1992), and large coronal and chromospheric
fluxes (Pakull 1981; Vilhu, Gustafsson, \& Walter 1991), AB~Dor
also possesses co-rotating material at 2$-$5 stellar radii
(Cameron \& Robinson 1989), in the form
of cool prominences or H$\alpha$ clouds (Cameron et al.\/ 1990).
Its brightness and activity levels make it an ideal target
to study the extremum of stellar magnetic activity.
Among the recent studies are those
by Rucinski et al.\/ (1995b); Mewe et al.\/ (1996);
K\"urster et al.\/ (1997); Schmitt, Cutispoto, \& Krautter (1998)
Vilhu et al.\/ (1998), and Ake et al.\/ (2000).

The parallax (Guirado et al.\/ 1997)
places the star at 15~pc, slightly above the zero-age main sequence
(Cameron \& Foing 1997), with a probable
age close to that of the Pleiades.
AB Dor is a member of a multiple star system. The dMe4 star
Rossiter~137B (AB~Dor~B) is a
common proper motion companion at an angular distance of 10~arcsec
(Innis et al.\/ 1985; Innis, Thompson, \& Coates 1986; Vilhu et al.\/ 1989). 
Guirado et al.\/ (1997) detected a low mass astrometric companion, AB~Dor~C, 
with an inferred separation of a few tenths of an arcsec. R~137B is outside
the GHRS aperture; AB~Dor~C falls within the aperture, but is unlikely to
make any significant contribution to the observed flux.

\subsection{The November 1994 Campaign}

We observed AB Dor with the Goddard High Resolution Spectrograph
(GHRS; Brandt et al.\/ 1994; Heap et al.\/ 1995; Robinson et al.\/ 1998) aboard
the Hubble Space Telescope, as part of a multiwavelength campaign
in November 1994 (Walter et al.\/ 1995). Our goal was to obtain simultaneous
spectroscopic and photometric observations at X-rays, UV, optical, and
radio wavelengths over at least one stellar rotation period, in order to
correlate the coronal, chromospheric, and optical behaviors, and to come up
with a 3-dimensional picture of the atmosphere of a very active star.

The optical spectroscopy and photometry, and the Doppler images,
have been reported by Cameron et al.\/ (1999). AB Dor behaved 
normally (for AB~Dor). There was a prominent photometric wave, indicating a 
highly asymmetric starspot distribution. We observed no large flares (i.e.,
flares with durations in excess of about an hour).
Cameron et al.\/ (1999) report a number of strong absorption events due to cool
material (extended prominences) co-rotating at high altitude. The Doppler
image (Cameron et al.\/ 1999) shows a dark feature at high latitudes, with
some low latitude spottedness. Vilhu et al.\/
(1998) report a continuous GHRS observation of the \ion{C}{4} line which
immediately preceeded our observations.

We will report on the variability of the chromosphere and corona in a
subsequent paper (Walter et al., in preparation). Here we present an
analysis of the quiescent chromosphere of AB Dor as viewed over one stellar
rotation period by the GHRS instrument on the HST.

\section{The Observing Program and Data Reductions}

We observed AB Dor (program 5181) with the HST/GHRS on 1994 November 14-15
while the target was in the continuous viewing zone.
The observations began near the end of stellar rotation cycle
10440\footnote{using the ephemeris HJD = 2444296.575 + 0.51479E}.
Our strategy was to maximize spectral and
temporal coverage by rapidly changing gratings to obtain a sequence of
exposures during the 12.3 hour rotation period of AB Dor. For 12 hours
we alternated between 256 second integrations of  the \ion{Mg}{2} lines,
using the Echelle-B (R$\equiv\frac{\lambda}{\delta\lambda}~\sim$~100,000), and
1228 second integrations of transition region lines
using the G160M first order grating (R$\sim$20,000).
This strategy was driven by the one dimensional format
of the GHRS detectors; at R=20,000 we observe only about 35\AA\ at a time.

The main observation was split into 6 sequences.
Each sequence began with a peakup in the large aperture, to ensure
that the target remained well-centered. Each sequence then consisted of
3 observations of the \ion{Mg}{2}
line alternating with observations of the \ion{Si}{4}
$\lambda\lambda$1393,1402 doublet, the \ion{C}{4}
$\lambda\lambda$1548,1551
doublet, and either the \ion{N}{5} $\lambda\lambda$1238,1242 doublet, 
the \ion{C}{2} $\lambda\lambda$1334,1335 doublet, or the
density-sensitive
\ion{Si}{3}], \ion{C}{3}] $\lambda\lambda$1892,1908 intersystem lines.
We concluded these six sequences with a final 
\ion{Si}{4} exposure sandwiched between two \ion{Mg}{2} observations.
All these observations utilized the D2 detector.
At the end of these sequences, we changed to the D1
detector and observed the full short-wavelength spectrum 
(1150-1750\AA) at low dispersion in two G140L exposures.
The details of the exposures are given in Table~\ref{tbl-obslog}.
Four observations were lost when the carrousel failed to lock.

\placetable{tbl-obslog}

Each science observation was preceeded by a spectral calibration lamp
exposure which established the accurate wavelength scale.
Occasional
SPYBAL observations of the wavelength calibration lamps were also interspersed
by the scheduling software.
The target was observed through the large science aperture (LSA)
using substep pattern 5. We calibrated
the raw data using the GHRS team software procedure
CALHRS (Blackwell et al.\/ 1993). Each \ion{Mg}{2}
observation is a single spectrum, but
 each transition region spectrum consists of 4 independent
spectra, providing about 5 minute temporal resolution. 
The analysis of the line profiles (excluding the \ion{Mg}{2} ISM analysis
in section~\ref{sec-mgism}) was carried out using the
ICUR software package\footnote{
\url{http://sbast3.ess.sunysb.edu/fwalter/ICUR/icur.html}}.
All the measurements
(fluxes, centroids, Gaussian fits) use standard techniques. 

Mewe et al.\/ (1996) fit the EUVE and ASCA spectra, and found a best fit
absorption column of
2.4$\pm$0.5~$\times$10$^{-18}$~cm$^{-2}$. Since the reddening correction
is less than 1\%, we do not apply any reddening corrections.

\section{The Low Dispersion Spectral Atlas}

\begin{figure}
\plotone{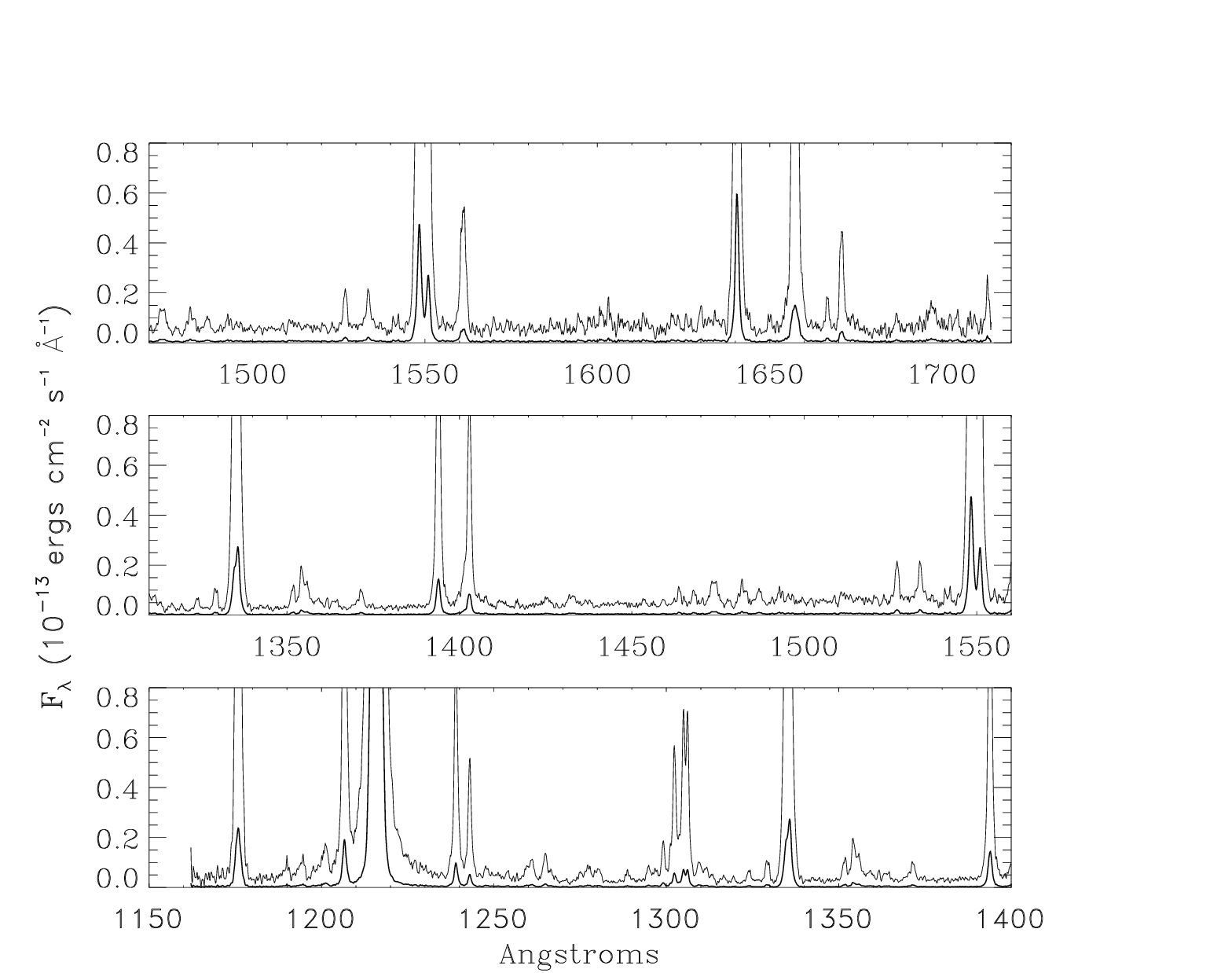}
\caption{The summed low dispersion G140L
spectrum of AB Dor between 1162 and 1714\protect\AA,
smoothed with a Fourier filter.
The data are binned into 0.14\protect\AA\ bins.
The flux scale refers to the
thin line; the thick line is scaled down by a factor of 10 to show the full
dynamic range in the spectrum.
Between 1428 and 1448\protect\AA, where the spectra overlap, the effective
exposure time is 1688~s; elsewhere the exposure time is 844~s. Much of the
Lyman~$\alpha$ emission is geocoronal. Geocoronal O~I \protect$\lambda$~1304
is seen as
a low broad pedestal under the narrow stellar lines.
}\label{fig-g140l}
\end{figure}

The summed low dispersion spectrum is shown in Figure~\ref{fig-g140l}.
Two spectra centered at 1305\AA\ and 1571\AA\ cover almost all
the useful wavelength range of the G140L grating (1162-1714\AA ) with
20\AA\ of overlap in the middle. The resolution of this spectrum, 
R$\sim$2000,
is adequate to resolve most astrophysically-important blends, including
(marginally) the \ion{C}{2}~$\lambda$1334,1335 doublet.
We obtained this spectrum
primarily for the purpose of constructing the line atlas
(Table~\ref{tbl-lines}), and to determine the emission measure distribution
(Section~\ref{sec-emd}) of the chromosphere of this very active star.

\placefigure{fig-g140l}
\placetable{tbl-lines}

The spectrum shows a wealth of detail, including weak lines of
neutral and low-ionization species, the strong transition region lines, and
the \ion{Fe}{21} coronal line.
We have identified the emission lines using the solar line lists of
Burton \& Ridgeley (1970)
and Feldman et~al.\/ (1997). Other possible identifications are found in 
the CHIANTI (Dere et al.\/ 1997) database.
The brightest stellar emission line is \ion{He}{2}~$\lambda$1640 (the
H~I~Lyman~$\alpha$ emission is geocoronal).
Identifications for some of the weaker lines
and blends are uncertain,
and will require higher dispersion spectra for definitive
identification.

There are 63 lines in Table~\ref{tbl-lines} with secure identifications.
The mean difference between the observed and tabulated (rest)
 wavelengths is 0.15$\pm$0.02\AA , corresponding to a radial velocity of 
33$\pm$5~km~s$^{-1}$, in good agreement with the previously-measured
photospheric value of
30~km~s$^{-1}$. Much of the scatter lies, as expected, in the weaker lines.
The mean radial velocity of the 34 strongest lines
(fluxes $>$1.0$\times$10$^{-14}$~erg~cm$^{-2}$~s$^{-1}$) is also
33$\pm$5~km~s$^{-1}$. This agreement with the photospheric radial velocity
verifies both that the wavelength scale is accurate and that the line 
identifications are most likely correct.

\section{The $\lambda$1900\AA\ Region\label{sec-1900}}

\begin{figure}
\plotone{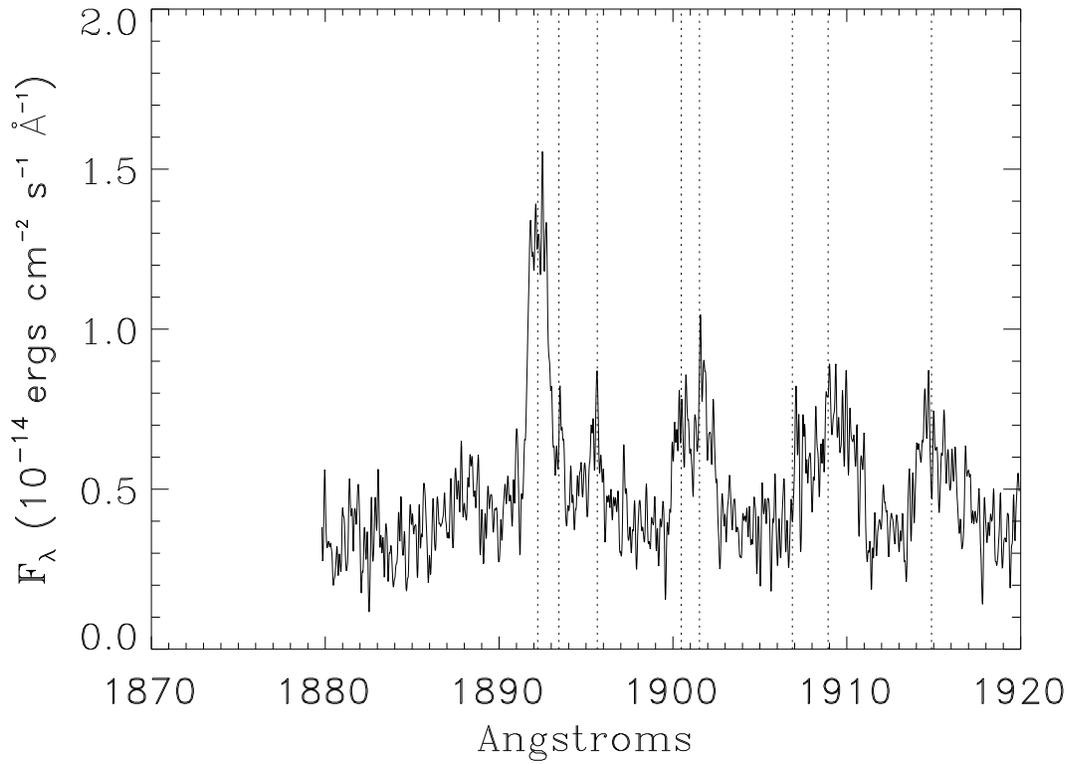}
\caption{The mean G160M spectrum of the 1900\AA\ region. The two observations
have been averaged, and a Fourier filter has been applied. In addition to the 
\protect\ion{Si}{3}]~$\lambda$1892.03 and 
\protect\ion{C}{3}]~$\lambda$1908.734 lines,
we see emission of \protect\ion{S}{1}~$\lambda\lambda$1893.252, 1895.459,
1900.27, and 1914.68 and
\protect\ion{Si}{1}~$\lambda$1901.338.
There appears to be a narrow emission line at 1907.34\AA.
The \protect\ion{C}{3}] $\lambda$1908.734 line is broader than the other lines,
perhaps because it arises in low density regions well above the photosphere
but co-rotating with the star.
The dotted vertical lines mark the rest wavelengths of these lines at the
stellar radial velocity.
}\label{fig-1900}
\end{figure}

Figure~\ref{fig-1900} shows the mean spectrum in the $\lambda$1900\AA\
region. The two G160M observations were obtained at
rotational phases 0.336 and 0.800, and sample opposite hemispheres of the
star. The $\lambda$1900\AA\ continuum is 46\% brighter at phase 0.8. 
The contemporaneous $U$ and $V$ band photometry (Rucinski, Garrison, \& 
Duffee 1995a; Cameron et al.\/ 1999) shows that the
star was faintest at about rotation phase 0.5, and that in the $U$~band AB~Dor
is about 15\% brighter at phase 0.8 than at phase 0.34. There is no
significant change in the emission line fluxes between the two observations.

\placefigure{fig-1900}

In addition to the expected
\ion{Si}{3}]~$\lambda$1892 and \ion{C}{3}]~$\lambda$1908 intersystem
lines, we see four narrow lines of \ion{S}{1} and a line of \ion{Si}{1}
(Table~\ref{tbl-1900}, Figure~\ref{fig-1900}).
The possible feature at 1907.3\AA\ (1907.1\AA\ rest wavelength) is
unidentified.
The \ion{C}{3}]~$\lambda$1908 line
is much broader than the other lines (see Section~\ref{sec-dens4}). The line
widths
listed in Table~\ref{tbl-1900} are measured by fitting Gaussians to the lines.
A  line broadened by the stellar rotation would have a FWHM of approximately
0.8\AA\ in this spectral region (assuming a 40\% overestimate of 
$v$~sin~$i$ from the Gaussian fit: see discussion in Section~\ref{sec-trp}).

\placetable{tbl-1900}

\section{Transition Region Quiescent Line Profiles\label{sec-trp}}

\begin{figure}
\plotone{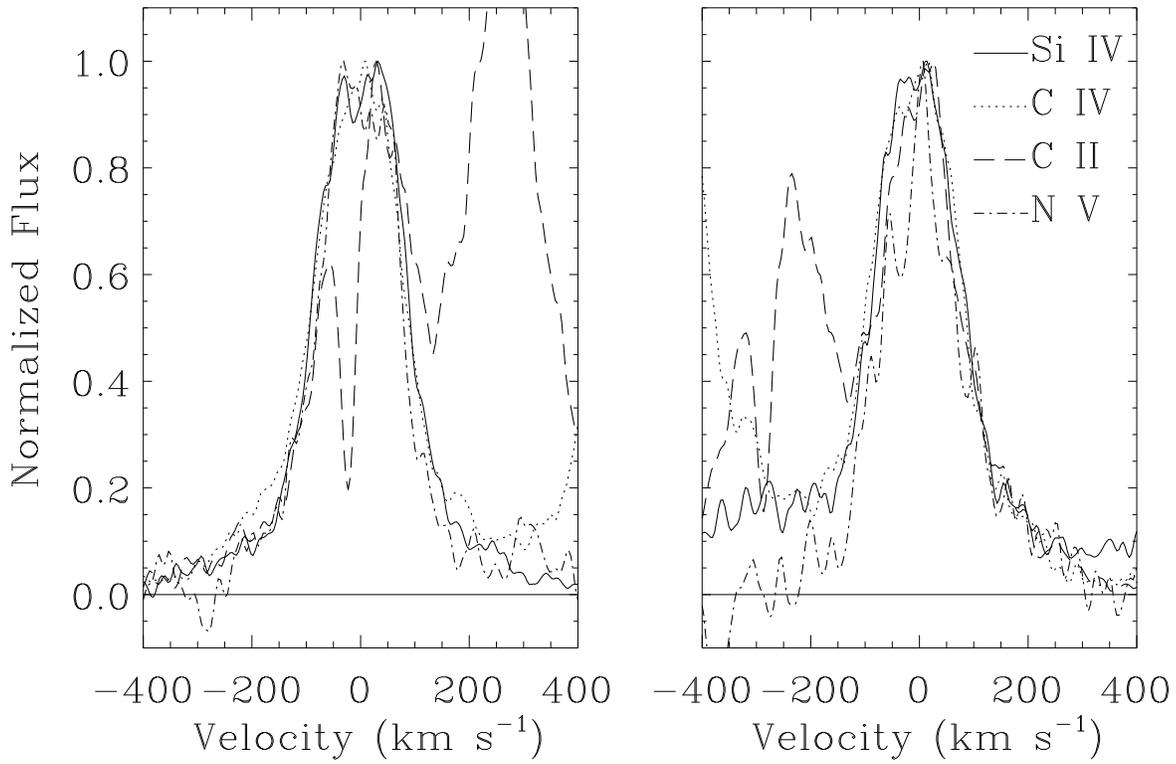}
\caption{A comparison of the 4 transition region doublets in velocity space.
The left panel shows the blueward line; the right panel shows the redward line.
The normalized profiles are similar, with the following exceptions: 
the \protect\ion{C}{2} $\lambda$1334 emission line includes
a strong interstellar absorption feature;
the blue wing of the \protect\ion{Si}{4} $\lambda$1402 line (right panel)
is elevated due to 
\protect\ion{O}{4} emission; and
the \protect\ion{N}{5} $\lambda$1242 line is noisy.
The other component of the \protect\ion{C}{2}
doublet is visible in each panel, and
the wings of the \protect\ion{C}{4} lines overlap.
}\label{fig-tr}
\end{figure}

We monitored the transition region doublets of
\ion{C}{2}, \ion{Si}{4}, \ion{C}{4}, and \ion{N}{5}. The 
\ion{Si}{4} and \ion{C}{4} lines were each observed six times, and
exhibited variability. 
Each observation consisted of four $\sim$~5 minute integrations, which 
permits us to examine the data for line profile variations or flaring on
timescales of 5 to 20 minutes. We examined the individual integrations for
evidence of flaring (rapid flux variations over the 4 integrations, 
varying line profiles, or fluxes significantly in excess of the median). 
We generated quiescent
line profiles by summing those spectra which had fluxes near the median,
symmetric line profiles, and no evidence for flaring. 
In Figure~\ref{fig-tr} we compare the 4 quiescent line profiles. 
Here we discuss only the quiescent line profiles;
we will discuss the flaring spectra elsewhere.

\placefigure{fig-tr}

We initially fit the emission lines as Gaussians or sums of Gaussians because
this approach is commonly used
and the resulting fits are generally quite acceptable.
Vilhu et al.\/ (1998) found that the \ion{C}{4} emission lines could
be fit as a sum of narrow and broad Gaussian components (as found in general 
by Wood, Linsky, \& Ayres 1997). The narrow components are significantly
broader than the expected rotational broadening, and Vilhu et al.\/ 
interpreted this nonthermal broadening in terms of solar-like nonthermal
broadening mechanisms (e.g., Dere \& Mason 1993),
perhaps related to microflares. Unlike the case of the Sun, we do not know
much about the spatial distribution of the emitting gas. It probably is not 
confined to a thin region, given the evidence for extended prominences, in
which case the atmosphere is extended and interpretation of line profiles
is not straightforward.

We simulated line profiles from photospheres (with limb darkening parameters 
$-$1.0~$<~\epsilon~<$1.0) and from optically-thin extended
co-rotating atmospheres,
convolved them with the expected thermal broadening and
with a Gaussian instrumental response, and fit the lines as Gaussians.
In the case of $\epsilon$=0.6 we recover the well-known rotational profile
(e.g., Gray 1992); $\epsilon$=0.0 is a uniform disk, and $\epsilon<$0 yields
limb-brightened profiles. Transition region lines are more likely
optically thin emission from an extended atmosphere.
We also examined cases where the emission is confined to low latitudes.
Limb-brightened atmospheres tend to produce flat-peaked line profiles
(for modest extents of $\la$10\% of the stellar radius),
whereas limb-brightened atmospheres with emission confined to
low latitudes produce emission cusps for sufficiently large $v$~sin~$i$.

Although the intrinsic line profiles are not Gaussian, significant noise and
low resolving power allow Gaussian fits to be acceptable in many cases.
The widths of the Gaussian fits always exceed $v$~sin~$i$.
We find that for photospheric line profiles, the Gaussian fits overestimate
$v~$sin~$i$ by up to 40\%, depending on the value of $\epsilon$ and the
intrinsic
$v~$sin~$i$. Gaussian fits to limb-brightened profiles
(containing a few thousand counts) overestimate the $v~$sin~$i$ by 40-80\%.
Dere \& Mason (1993) showed that optical thickness will also tend to
broaden the line profiles; in slightly thick but effectively thin lines,
photons from the line core will be scattered into the wings.
We caution the reader that the interpretation of emission line profiles is
highly model dependent. Consequently, we do not interpret the width of the
narrow emission component of the transition region lines.

\begin{figure}
\plotone{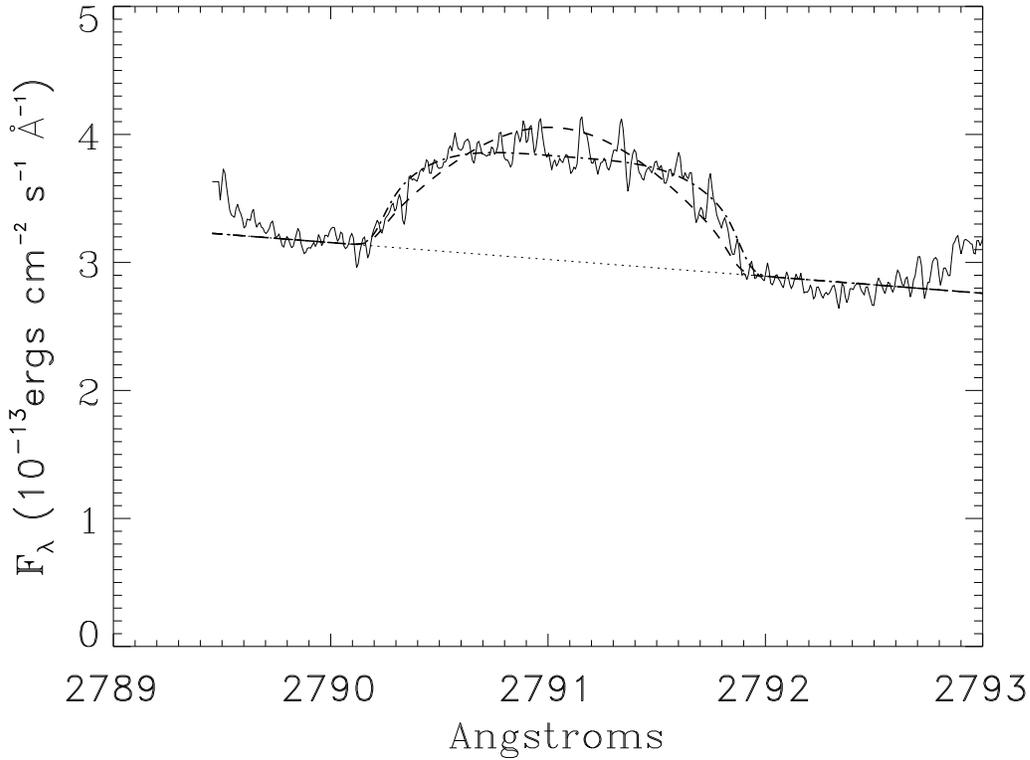}
\caption{The subordinate $\lambda$2791 line of \protect\ion{Mg}{2}. The data are
the mean of 19 echelle observations.
This line is almost certainly optically thin, and therefore
represents the convolution
of the surface distribution of \protect\ion{Mg}{2}
with the rotational and thermal
broadening. We show two fits to this line. The narrower profile (dashed)
is the rotational profile for a limb-darkened surface ($\epsilon$=0.6); the
broader profile (dash-dot), which better matches the data, is a limb-brightened
atmosphere with a height of 2\% of the stellar radius. Both profiles are
rotationally broadened with $v$~sin~$i$=90~km~s$^{-1}$, and have 5~km~s$^{-1}$
of thermal broadening. There is no evidence for non-thermal broadening at
this level of the atmosphere.
The dotted line is the extrapolated continuum.
}\label{fig-mg2s}
\end{figure}

In Figure~\ref{fig-mg2s} we show the subordinate
(3p $^2$P$_\frac{1}{2} -$ 3d $^2$D$_\frac{3}{2}$) \ion{Mg}{2} line at
$\lambda$~2791.6\AA , which is almost certainly optically thin. The
profile is flat-topped, and at resolution R$\sim$10$^5$ the shape of the
line profile is clearly not Gaussian.
The rotation profile for a limb-darkened photosphere is much more sharply
peaked. Figure~\ref{fig-mg2s} demonstrates that this line can be produced by a
limb-brightened, optically thin atmosphere with a height of a few percent of
the stellar radius. This fit is not unique: within the uncertainties
one could also fairly well reproduce the profile with, for example, a
rotating photospheric profile ($\epsilon$=0.6) with no emission above about
30$^\circ$ latitude, or a uniform disk profile ($\epsilon$=0.0) with emission
confined below about 60$^\circ$ latitude.

\placefigure{fig-mg2s}

The strong resonance lines of the transition region
clearly cannot be fit with a single component. Instead we fit
the lines as sums of narrow and broad Gaussians. We find that we get better
fits with three components: a broad component and two similar narrow
Gaussians displaced from line center by about $\pm$~0.5~$v$~sin~$i$. This
model gives better fits because the peaks of the \ion{Si}{4} and \ion{C}{4} are
flatter than a Gaussian profile. Physically, this is because in an
extended limb brightened atmosphere the centroid of the emission is displaced
to higher velocities where the path length through the atmosphere is longer.
We also have had success in fitting the line wings with emission from
an optically thin, extended, rigidly-rotating atmosphere (see
Section~\ref{sec-trd}).

\subsection{\ion{C}{4}}

We identified 17 of the 24 \ion{C}{4} integrations as likely to represent the
non-flaring, quiescent transition region.
The \ion{C}{4} flux (the sum of the 
fluxes in the $\lambda\lambda$1548,1551 lines) during quiescence
is less than 1.5$\times$10$^{-12}$~erg~cm$^{-2}$~s$^{-1}$. 
Vilhu et al.\/ (1998) noted that the non-flaring flux level just prior to our
observation ranged between
1.0 and 1.5$\times$10$^{-12}$~erg~cm$^{-2}$~s$^{-1}$, and the
average \ion{C}{4} flux during our entire observation is 
1.2$\times$10$^{-12}$~erg~cm$^{-2}$~s$^{-1}$.
The $\lambda\lambda$1549/1551 line ratio of 1.78$\pm$0.03 is close to
but less than the optically thin limit.

\begin{figure}
\plotone{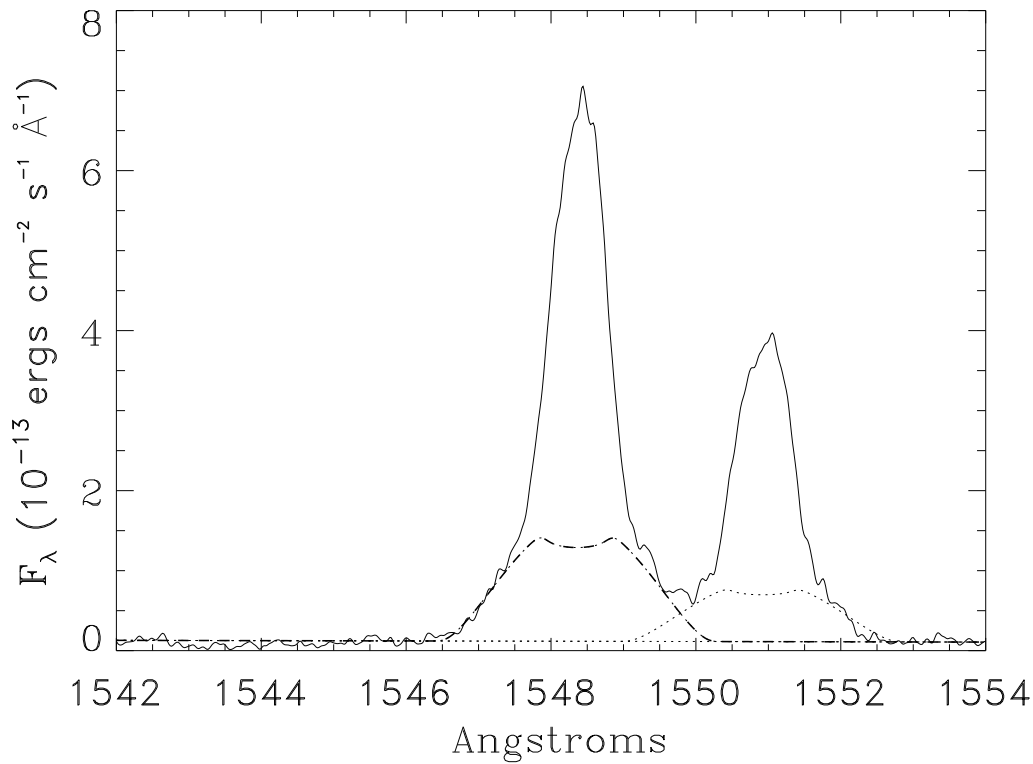}
\caption{The mean quiescent \protect\ion{C}{4} line profiles. The data have been
smoothed with a Fourier filter.
The dashed and dotted lines represent a model for emission from an
optically-thin, constant density atmosphere extending out to 3 stellar radii.
The model is centered on the rest wavelength of the line, and is scaled to the
blue wing.
The $\lambda$1551\AA\ line is not fit, but is assumed to be
half the intensity of the $\lambda$1548\AA\ line. 
}\label{fig-civ}
\end{figure}

The summed quiescent line profile (Figure~\ref{fig-civ})
should represent the phase-averaged transition region. 
Although we can fit the line with a sum of two Gaussians with fit
parameters similar to those of Vilhu et al.\/ (1998), we find that the data
are better fit with three components. 
We plot neither the 2- or 3-component Gaussian fits to this or any
line in Figures~\ref{fig-civ} to \ref{fig-mg2k}. Rather, 
the fits to the high velocity wings shown in Figure~\ref{fig-civ} are for
optically thin, constant density atmospheres extending out 2.5 stellar
radii. Forty percent of the total flux is in the extended component.

\placefigure{fig-civ}

\subsection{\ion{Si}{4}}

\begin{figure}
\plotone{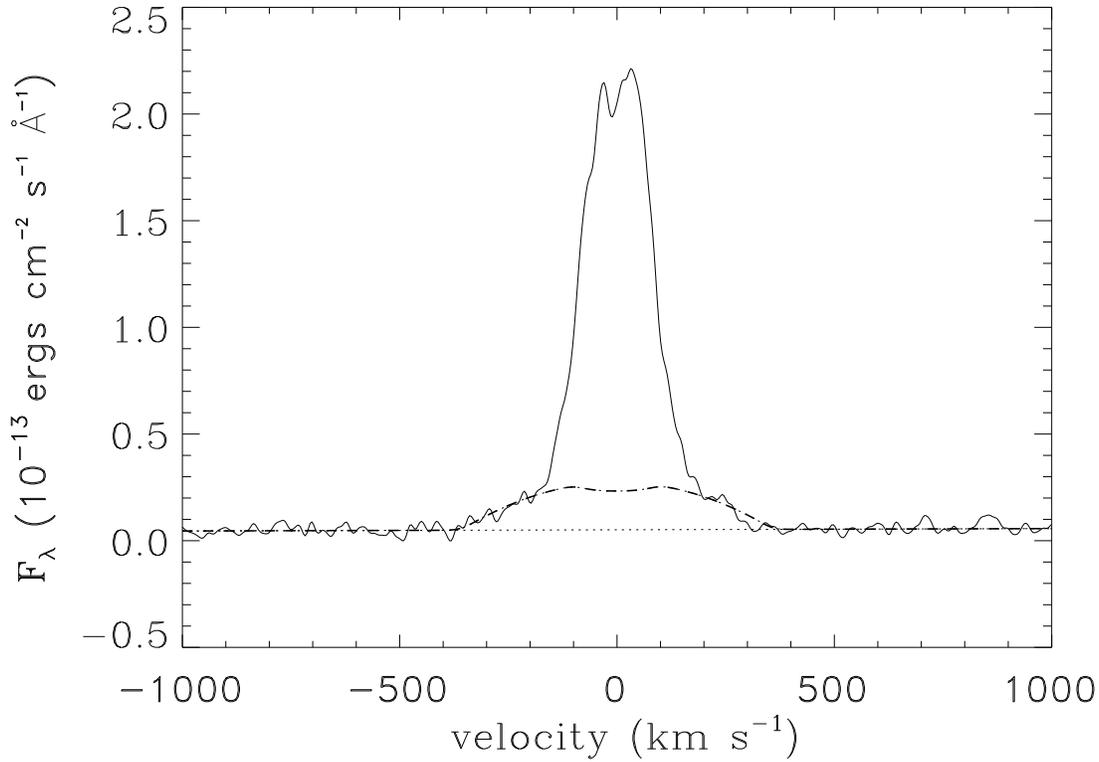}
\caption{The mean quiescent \protect\ion{Si}{4} $\lambda$1393 line profile.
The data have been smoothed with a Fourier filter.
The dashed line represents a model for emission from an
optically-thin, constant density atmosphere extending out to 3 stellar radii.
The double-peaked line profile is not representative of any individual line,
but is the consequence of summing 20 variable line profiles.
}\label{fig-siiv}
\end{figure}

The mean \ion{Si}{4}~$\lambda$1393 line flux is
2.2$\times$10$^{-13}$~erg~cm$^{-2}$~s$^{-1}$. From a
visual examination of the line profiles and the line fluxes, we identified
20 quiescent spectra, and summed these (Figure~\ref{fig-siiv}).
The $\lambda\lambda$1393/1402 line ratio of 1.78$\pm$0.02 is identical to
that in the \ion{C}{4} lines. Although close to the optically thin limit,
both line ratios are significantly less than 2. This probably indicates that at
least some of the emission is from optically thick regions.

Because the lines are well separated and because the blue wing of the
$\lambda$1402 line is blended with \ion{O}{4}] $\lambda$1401.34, we
fit only the $\lambda$1393 line. 
As with the \ion{C}{4} lines, the \ion{Si}{4} line profile is better fit
with a broad component 
centered on the rest velocity of the star plus two narrow emission
components than it is with a two-Gaussian model. If we fit the high velocity
wings as emission from extended gas (see Figure~\ref{fig-siiv}), we find
that the emission extends out to
3 stellar radii, which is comparable to what we found for \ion{C}{4}.
About 25\% of the \ion{Si}{4} emission is in the extended component.

\placefigure{fig-siiv}

\subsection{\ion{N}{5}}

We obtained a single G160M observation of the \ion{N}{5} doublet at phase
0.5. The line fluxes are 95\% of those seen in the low dispersion spectrum.
The line ratio of 2.10$\pm$0.04 is consistent with the line's being
optically thin. There is no evidence for the broad component seen in
the \ion{Si}{4} and \ion{C}{4} lines, perhaps because of the low S/N in this
integration. However, there is a small flux excess on the red side of the 
$\lambda$1242\AA\ line. This is not seen on the blue side of the line, nor
is it obvious in the $\lambda$1238\AA\ line.
The cause could be a downflow, or perhaps a small flare on the receding
limb of the star.


\subsection{\ion{C}{2}}

We obtained a single G160M observation of the \ion{C}{2} doublet at phase
0.65. The emission lines are severely blended,
and we were unable to 
fit the various components uniquely. High velocity wings are evident. There is
a strong interstellar absorption feature in the $\lambda$1334.5\AA\
line, with a radial
velocity of 9$\pm$3~km~s$^{-1}$. Given the resolution of
the GHRS in this mode, it is not possible to identify the absorption with any
particular component of the interstellar medium (see Section~\ref{sec-mgism}).
Corresponding (though much weaker) absorption is seen near the top of the
$\lambda$1335.7\AA\ line.

%

This spectrum also shows the \ion{S}{1} $\lambda$1323.52, 
\ion{C}{1} $\lambda$1329.1, and \ion{Cl}{1} $\lambda$1351.66 lines with fluxes
similar to those in Table~\ref{tbl-lines}.

\subsection{Discussion of the Transition Region Lines\label{sec-trd}}

The transition region lines of active stars are generally modeled as the sum of 
broad and narrow Gaussians. 
The width of the narrow component is generally significantly
broader than the stellar $v$~sin~$i$, and this is interpreted as the
convolution of the stellar rotational profile with some other broadening
mechanism. Wood, Linsky, \& Ayres (1997) showed that among dwarf
stars, the excess broadening appears to decrease with increasing surface
gravity ($\xi_{NC}~\propto~g^{-0.68}$).
The broad component is often interpreted as evidence for high velocity gas
associated with microflaring. 

We have difficulty interpreting the line profiles of AB Dor in this manner. 
After deconvolving the 90~km~s$^{-1}$ rotational velocity from the narrow
central emission component we find that the velocity of the excess
broadening mechanism is about 120~km~s$^{-1}$, which places AB~Dor
well above the trend found by Wood, Linsky, \& Ayres (1997) for less active
stars. And while microflaring
may indeed exist and contribute to the width of the broad component,
the evidence
for spatially extended gas leads us to propose a fundamentally different
mechanism for the broad line component.

We propose that the high velocity wings are due to spatially-extended
gas co-rotating with the star,
and not to some unknown broadening mechanism. The existence of
extended prominences seen in the light of H$\alpha$ is well established;
it is likely that this
same gas accounts for the extended wings of the \ion{Mg}{2} $k \& h$ lines.
The wings of the transition region lines have very similar shapes, suggesting
that these lines also have
a similar origin. Corroborating evidence for extended hot gas hot gas
exists: Walter et al.\/ (1995; 2001 in preparation) show that C~IV undergoes
absorption events associated with H$\alpha$ prominences. 

We model the extended gas as a simple optically thin volume co-rotating with
the star, and assume that the density is uniform between some inner
and outer radii. 
The emission from each point is broadened by a thermal velocity appropriate to
the ion in question. The computed profiles do not depend sensitively on the
value of the assumed inner radius, since the
emission from gas close to the star is hidden by the central emission
component, but the outer radius does determine the width of the broad line
component. We vary only the outer radius to match the observations. We can also
vary the extent of the gas in latitude, but we are not very sensitive to
this because gas at high latitudes has smaller velocities, and is buried
under the central emission.

The inferred radial extent of the extended gas is close to the Keplerian
co-rotation radius of 2.8 stellar radii. We favor a model in which the gas is
trapped by large-scale quasi-dipolar magnetic fields (Walter 1999). Such gas
will be most stable near the co-rotation radius (e.g., the ``slingshot
prominences'' of Cameron \& Robinson 1989). 

Assuming that the gas is optically thin and in rigid rotation, 
we can use the high velocity wings to estimate the mean electron density
in the extended gas. The mean electron density n$_e$ is given by
0.8~$\sqrt{\frac{L}{PV}}$, where L is the velocity-resolved luminosity of the
line, P is the power in the line, V is the velocity-resolved volume
element, and the 0.8 factor accounts for the fact that hydrogen is largely
ionized.
The mean density is about 5$\times$10$^7$~cm$^{-3}$ at the temperature
of the transition region lines
for a filling factor of 1; the presence of
discrete prominences suggests that the filling factor of this extended region
is very low, and the density in the prominences must be commensurately larger.

We see no evidence for any rotational modulation of the fluxes in the
\ion{Mg}{2}, \ion{C}{4}, or \ion{Si}{4} lines.   
The photosphere is faintest at rotation phase 0.5 (Cameron et al.\/ 1999).
Following solar analogy, the dark regions (starspots) are likely regions of
larger photospheric magnetic flux and the bright chromosphere and
transition region should be located above the starspots.
If the chromosphere and transition region are spatially
extended, as is likely the case here, then any modulation would be diluted.
Our data let us place a limit of 5\% on the amplitude of the
rotational modulation in the chromospheric and transition region lines.

\section{The \ion{Mg}{2} lines}

\subsection{The Chromospheric \ion{Mg}{2} Emission}

\begin{figure}
\plotone{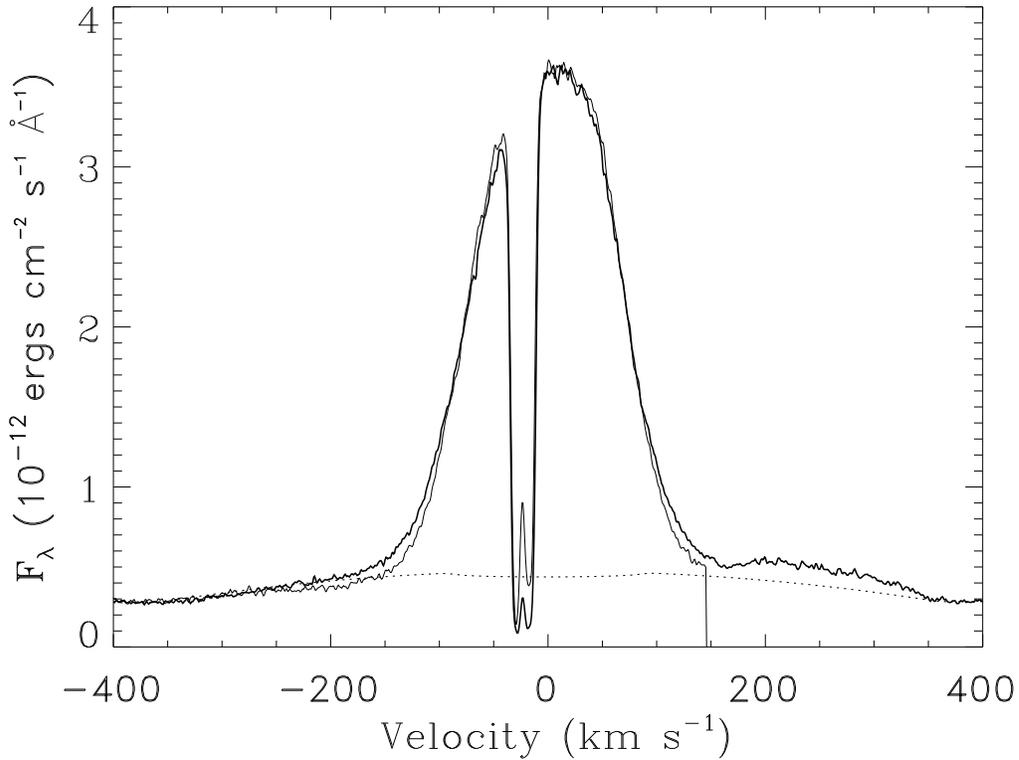}
\caption{The mean \protect\ion{Mg}{2} line profiles.
The thick line is the $k$ line;
the $h$ line is overplotted as the thin line (the data are cut off at 
+145~km~s$^{-1}$ by the edge
of the detector). The $h$ line flux has been multiplied by
a factor of 1.23 to match the line cores. The $k$ line is brighter in the wings.
The blue wing extends out to about -320~km~s$^{-1}$. The 
hump on the red wing of the $k$ line (between +170 and +350~km~s$^{-1}$)
is the sum of the
$\lambda$~2797.922 (3p $^2$P$_\frac{3}{2}$ -- 3d $^2$D$_\frac{3}{2}$) and 
$\lambda$~2797.989 (3p $^2$P$_\frac{3}{2}$ -- 3d $^2$D$_\frac{5}{2}$) 
subordinate lines of \protect\ion{Mg}{2}. The dotted line shows the expected
emission line profile for an optically-thin uniform density
extended atmosphere with a height of 3 stellar radii.
}\label{fig-mg2k}
\end{figure}

We observed the chromospheric \ion{Mg}{2} $k$ \& $h$ lines 19 times,
approximately every 35-50 minutes, with a resolving power of about 10$^5$.
The mean profile of the two lines is presented in Figure~\ref{fig-mg2k}.
The $k$ \& $h$ resonance lines are well-exposed, and the \ion{Mg}{2}
subordinate
lines are detected as well. Both the $k$ \& $h$ lines exhibit extended blue
wings, to velocities of about -320~km~s$^{-1}$ with respect to the star. The
red wings cannot be measured because the $k$
line is blended with the subordinate
lines, and the red wing of the $h$ line falls off the edge of the detector.
The brightness ratio near line center is 1.23, confirming that the line cores
are optically thick, and the $k$ line has brighter 
line wings as expected for the higher opacity line. 
Over 10\% of the flux in the line core is lost to the narrow interstellar
absorption lines (cf.\ Section~\ref{sec-mgism}).

\placefigure{fig-mg2k}

Since \ion{Mg}{2} is formed in the chromosphere,
at about the same temperature as H$\alpha$, one
might expect to see absorption events in \ion{Mg}{2} similar to those seen in
H$\alpha$ (Cameron et al.\/ 1990, 1999), and indeed we do see such absorption
events. These will be discussed elsewhere (Walter et al.\/ 2001, in
preparation). The same prominences seen in
absorption against the stellar disk should be seen in emission off the limb, and
may account for the broad wings of the $k$ \& $h$ lines.
The maximum velocity is consistent with emission from gas in co-rotation at
heights of up to
3.5 stellar radii. In Figure~\ref{fig-mg2k} we overplot the expected emission
line profile, scaled to the data, for a co-rotating
atmosphere extending
to a height of 3 stellar radii. This is neither a fit to the data, nor it is
a unique description of the data, but it demonstrates that the extended
wings could arise from an extended atmosphere without assuming non-thermal
broadening mechanisms. This extended emission accounts for 20\% of the
total \ion{Mg}{2} $k$ \& $h$ emission, after correcting for the interstellar
absorption. The mean density in the extended atmosphere at chromospheric
temperatures is
about 3$\times$10$^8$~cm$^{-3}$ (for a filling factor of 1). This is about a
factor of 6 higher than the density inferred from the transition region lines,
and within the uncertainties in consistent with a constant pressure atmosphere.

\section{The Density Diagnostics}\label{sec-dens}

The mean density at a given temperature provides clues
to the gross morphology of the stellar atmosphere, since the emission measure
($\int$n$_e$n$_H$~dV) depends on both the density and the volume.
Even at very low spatial
resolution, the observed
density can help us discriminate between compact high density emitting regions
and large extended structures.
There are a number of density diagnostics in this region of the spectrum
(e.g., Doschek et al.\/ 1978). We use the line strengths in
Tables~\ref{tbl-lines}, \ref{tbl-1900}, and \ref{tbl-1400} to compute the
density sensitive line ratios, and use the CHIANTI software
package to convert the
line ratios to densities.

\placetable{tbl-1400}

\subsection{Densities involving the \ion{Si}{3} lines}

We used 3 \ion{Si}{3} lines and the sum of the $\lambda\lambda$1294-1301
lines to form 4 line ratios. These consistently yield densities between
2 and 3$\times$10$^{12}$~cm$^{-3}$ at a temperature of about 3$\times$10$^4$K
(Table~\ref{tbl-dens}).

\placetable{tbl-dens}

\subsection{Densities involving the \ion{C}{3} lines\label{sec-dens2}}

Cook \& Nicolas (1979) discuss using the \ion{C}{3} lines to determine
densities. 
We use the $\lambda$1175, 1247, and 1908\AA\ lines in this analysis.
The densities are not mutually consistent (Table~\ref{tbl-dens}).
Cook \& Nicolas (1979) note that in the
Sun the density ratios involving the $\lambda$1175 line are consistently off,
perhaps because the $\lambda$1175 line intensity is reduced by optical depth
effects. An increase of about a factor of two in the $\lambda$1175 line flux
would bring the $\frac{\lambda 1175}{\lambda 1908}$ density into agreement
with the $\frac{\lambda 1247}{\lambda 1908}$ density, at about 
10$^{11}$~cm$^{-3}$. The $\frac{\lambda 1175}{\lambda 1247}$ density would then 
become about 5$\times$10$^{11}$~cm$^{-3}$.

Using the $\frac{1175}{977}$ line ratio, Schmitt, Cutispoto, \& Krautter
(1998) and Ake et al.\/ (2000) both infer densities at the high density
limit, $>$10$^{11}$~cm$^{-3}$, which is consistent with our estimates.

The densities inferred from the \ion{C}{3} lines (formed at 6$\times$10$^4$K)
are smaller than those inferred from \ion{Si}{3}
(formed at about 3$\times$10$^4$K):
in a constant pressure atmosphere they should have
about half the density indicated by the \ion{Si}{3} lines.
This would be consistent with a model in which the bulk of the
transition region emission arises in constant pressure loops.

Cook \& Nicolas (1979) also estimate densities from ratios of \ion{Si}{3}]
$\lambda$1892 and \ion{Si}{4}~$\lambda$1402 to \ion{C}{3}]~$\lambda$1908.
These densities are valid so long as the relative abundances are solar.
The $\frac{\lambda 1892}{\lambda 1908}$ ratio has traditionally been used
as a density diagnostic in IUE spectra (Doschek et al.\/ 1978). 
We find this ratio to be about 1.3, which suggests a
density near 3$\times$10$^{10}$~cm$^{-3}$, but the
$\frac{\lambda 1402}{\lambda 1908}$
ratio implies a density nearer 3$\times$10$^{11}$~cm$^{-3}$,
using Cook \& Nicolas' model atmosphere ratios. Note that the \ion{Si}{3}]
$\lambda 1892$ and \ion{C}{3}]~$\lambda$1908 line profiles are very different
(Figure~\ref{fig-1900}), with the \ion{C}{3}] line nearly twice as broad,
suggesting that they might not be formed co-spatially.

\subsection{Densities from the $\lambda$1400\AA\ region}

\begin{figure}
\plotone{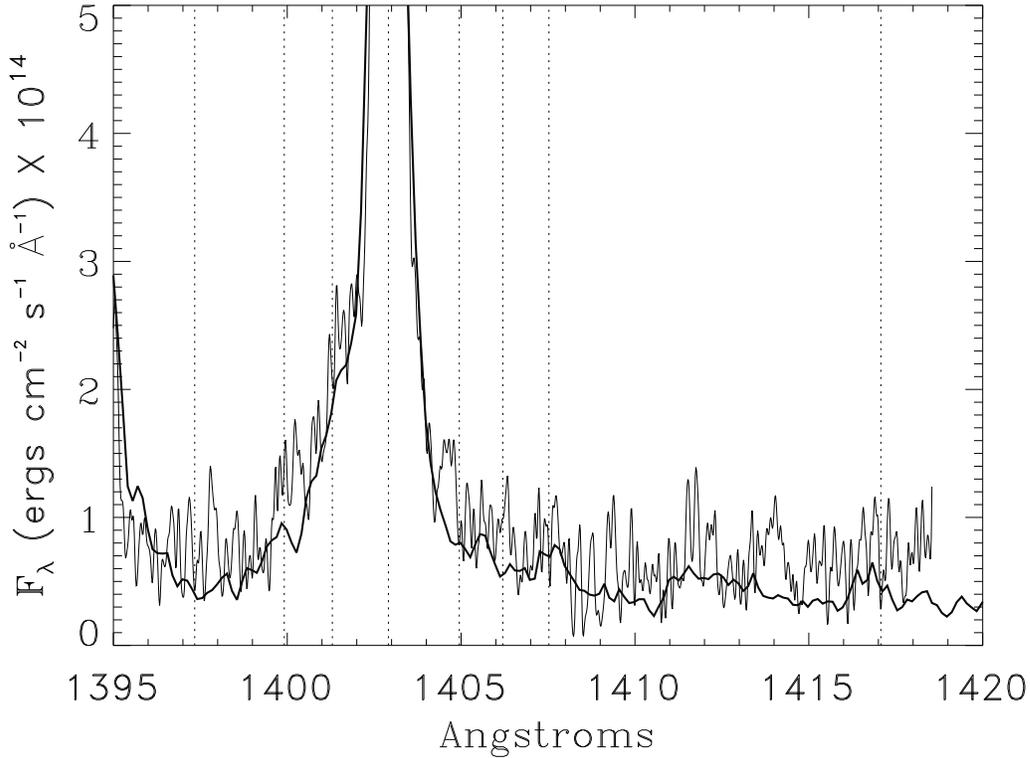}
\caption{Spectra in the 1400\AA\ region. The thin (noisier) line is the 
mean of 6 G160M observations (7368 seconds integration time);
the thick line is the 844 second G140L
spectrum. Both spectra have been smoothed with a Fourier filter. The strong
emission line is \protect\ion{Si}{4}~$\lambda$1402.
Vertical dotted lines mark the expected
locations of the \protect\ion{O}{4}] and \protect\ion{S}{4}]
lines, at the +30~km~s$^{-1}$
radial velocity of AB~Dor. The clear asymmetry in the blue wing of the
\protect\ion{Si}{4}~$\lambda$1402 line is due to the 
\protect\ion{O}{4}]
$\lambda$1399.8 and $\lambda$1401.2 lines. There may also be weak detections of
\protect\ion{O}{4}]~$\lambda$1407.4 and
\protect\ion{S}{4}]~$\lambda$1416.9 in the low
dispersion spectrum. The peak at $\lambda$1404.6 in the G160M spectrum
is at the wrong wavelength
to be the $\lambda$1404.8\AA\ \protect\ion{O}{4}]+\protect\ion{S}{4}] blend.
The other \protect\ion{O}{4}] and \protect\ion{S}{4}]
lines are not detected in either spectrum.
}\label{fig-1400}
\end{figure}

The $\lambda$1400\AA\ region (Figure~\ref{fig-1400})
includes a number of density-sensitive
\ion{O}{4}] (Cook et al.\/ 1995) and \ion{S}{4}]
(Dufton et al.\/ 1982) intercombination lines. These lines are much weaker
than the nearby \ion{Si}{4} resonance lines. The $\lambda$1399.8 and
$\lambda$1401.2 \ion{O}{4}] lines are clearly visible in
the low dispersion spectrum,
blended with the blue wing of \ion{Si}{4}~$\lambda$1402.8. We estimated their
fluxes first by fitting all 3 lines, forcing the wavelengths and line widths
of the \ion{O}{4}] lines to their expected values. 
Since the two \ion{Si}{4} lines should have identical profiles, 
we then shifted, scaled,
and subtracted the \ion{Si}{4}~$\lambda$1393.8 from the
\ion{Si}{4}~$\lambda$1402.8 line profile, and fit the residuals. The
$\lambda$1407.4 \ion{O}{4}] line is also visible in the low dispersion
spectrum. These fluxes can be found in Table~\ref{tbl-lines}.

\placefigure{fig-1400}

We also searched for these lines in the summed G160M spectrum
(total exposure time = 7368s). This spectrum is a global average over
all rotational phases. Although flaring affects
the strong \ion{Si}{4} resonance line profiles and fluxes,
we find no evidence of enhanced emission in the weaker lines during
flares. As in the low dispersion
image, there is clear evidence for extra emission on the blue wing of the
\ion{Si}{4}~$\lambda$1402.8 line. 
Using the same multi-line fitting procedure, we measure the line fluxes given
in Table~\ref{tbl-1400}.
We note that high dispersion does not improve the flux measurements
signficantly, because the
lines are over-resolved due to rotational line broadening, and the 
rotational line broadening causes line blending.

We measure the $\frac{\lambda 1399.8}{\lambda 1407.4}$ line ratio to be 
2.1$\pm$0.5, which is marginally consistent with the expected value of 0.993. 
Due to the poor accuracy of the flux determinations, the density
diagnositics are inconclusive. Similarly, the \ion{S}{4} line ratios are
inconclusive. 
The \ion{S}{4} $\lambda$1423.9 line is blended at low resolution with
a stronger \ion{S}{1} line.
Although there is a weak line near $\lambda$1416.9 in the low dispersion
spectrum, no line of similar strength is seen in the higher resolution
spectrum. 

\subsection{The Densities of AB Dor\label{sec-dens4}}

The densities listed in Table~\ref{tbl-dens} range from about
10$^{9}$~cm$^{-3}$ to 3$\times$10$^{12}$~cm$^{-3}$. The densities 
derived from the four line ratios using the \ion{Si}{3} lines are
self-consistent but larger than the other diagnostics,
while those involving 
\ion{C}{3}]~$\lambda$1908 are systematically low. As discussed above,
an arbitrary doubling of the \ion{C}{3}~$\lambda$1175 line flux (justified
because it appears to works for the Sun)
brings the $\frac{\lambda 1175}{\lambda 1247}$
density into better agreement with the \ion{Si}{3} densities
(Table~\ref{tbl-dens}),
while leaving all the densities based on \ion{C}{3}]~$\lambda$1908
consistently lower by about an order of magnitude.

The systematically lower densities indicated by the ratios involving the
\ion{C}{3}]~$\lambda$1908 line can be reconciled if the
\ion{C}{3}]~$\lambda$1908 line is preferentially formed in the extended, low
density region near the co-rotation radius. This would explain both the
significant broadening of the \ion{C}{3}]~$\lambda$1908 line, and the enhanced
line strength. The FWHM of the \ion{C}{3}]~$\lambda$1908 line
(Table~\ref{tbl-1900}) corresponds to about
3~$v$~sin~$i$, and so is consistent with formation
in a region close to the co-rotation radius.
We believe that the most reliable density estimate is 
n$_e$=2--3$\times$10$^{12}$~cm$^{-3}$ from the \ion{Si}{3} line ratios
because of the consistency among the four ratios. The electron pressure at
30,000K would then be P$_e$=n$_e$T=6--9$\times$10$^{16}$~cm$^{-3}$K, about
100 times the value for the quiet Sun.

\section{The Emission Measure Distribution\label{sec-emd}}
\begin{figure}
\plotone{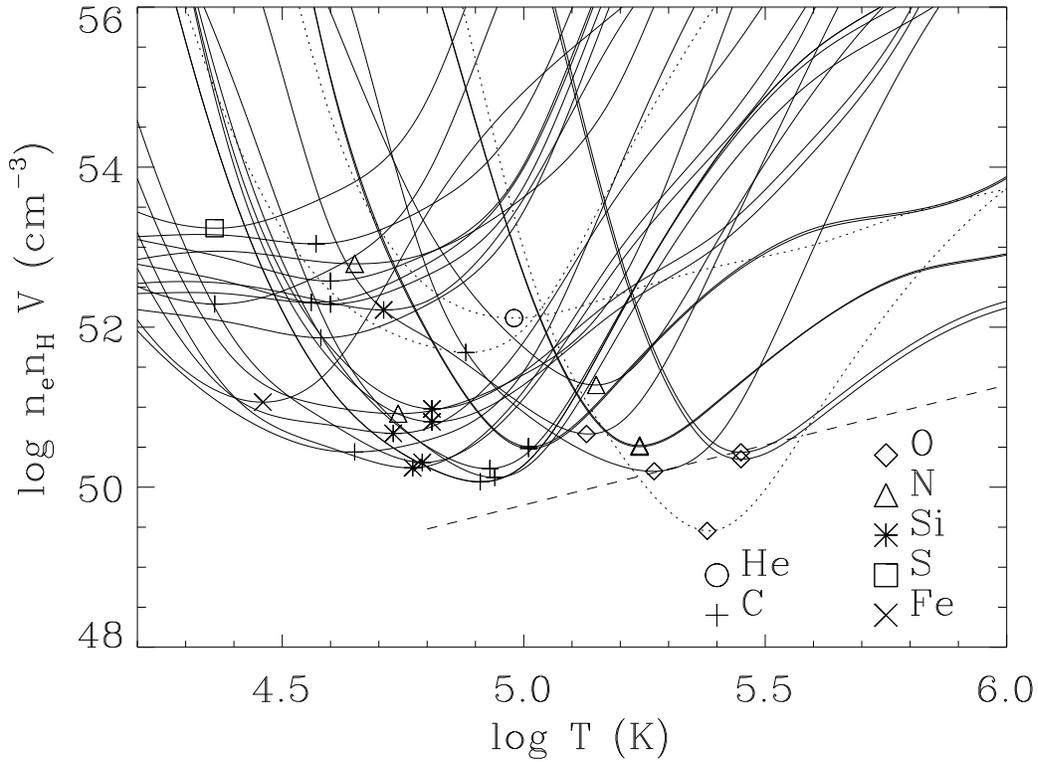}
\caption{The emission measure diagram for AB Dor.
Emissivities are from Landini \& Monsignori Fossi (1990), 
for an assumed electron density of 10$^{12}$~cm$^{-3}$. 
The emissivity curves have been interpolated with a cubic spline.
FUV data from Schmitt, Cutispoto, \& Krautter (1997) are included, as is the
T$^\frac{3}{2}$ extrapolation of the coronal emission measure from Rucinski
et al.\/ (1995b).
Symbols identifying the elements are plotted at the minimum emission measures
for each curve. The \protect\ion{O}{5} $\lambda$1371, \protect\ion{He}{2}
$\lambda$1640, and
\protect\ion{C}{3}] $\lambda$1908
emission measures, which are explicitly discussed in
section~\ref{sec-emd} of the text, are shown as dotted lines.
}\label{fig-emd}
\end{figure}

We present an emission measure distribution based on the observed
line fluxes in
Figure~\ref{fig-emd}. We use the line emissivities tabulated by 
Landini \& Monsignori Fossi (1990), interpolated with a cubic spline. 
Landini \& Monsignori Fossi computed the line emissivities in the low density
limit. Since much of the flux seems to arise in a high density atmosphere,
we used the CHIANTI database to determine the
relative emissivities at low and high densities, and used this to determine
proper emissivities for the few density-sensitive lines. 
The emission measure curves in Figure~\ref{fig-emd}
assume n$_e$=10$^{12}$~cm$^{-3}$. 

\placefigure{fig-emd}

In this diagram, 
we have overplotted the T$^\frac{3}{2}$ extrapolation of the coronal
emission measure, for $\int$n$_e$n$_H$dV=3$\times$10$^{52}$~cm$^{-3}$ at
log~T~=~6.6 (Rucinski et al.\/ 1995b; Mewe et al.\/ 1996). 
We have also added
the far UV line fluxes from ORFEUS II (Schmitt, Cutispoto, \& Krautter 1998),
after multiplying the ORFEUS II fluxes by 1.1 to bring their
\ion{C}{3}~$\lambda$1175 and
\ion{Si}{3}~$\lambda$1206 fluxes into agreement with our measurements.
The extrapolation of the emission measure of the cooler coronal component
is in good
agreement with the fluxes of the \ion{C}{3}, \ion{C}{4},
\ion{N}{5}, and \ion{O}{6} lines, especially considering the
non-simultaneity of the UV, FUV, and X-ray observations. 

A number of lines seem to have emission measures above the minimum.
The cool C, Si, and S lines may either form at temperatures below log(T)=4.2,
or they may derive much of their flux from a large extended volume.
Three other lines are clearly out of place.
\begin{itemize}
\item \ion{C}{3}]~$\lambda$1908 implies an emission measure 1.5 orders of
      magnitude larger than that of the other \ion{C}{3} lines, which are
      consistent with the trends from other species. No more than
      a few percent of the \ion{C}{3}]~$\lambda$1908 flux can be generated at
      these high densities due to strong collisional de-excitation;
      the remainder of the flux may arise in an extended lower density region.
\item \ion{He}{2}~$\lambda$1640 is far too strong to be formed in collisional
      equilibrium at a temperature near 10$^5$~K (Jordan 1975), given the
      emission measure at this temperature.
      This line is more likely formed during
      recombination following ionization of \ion{He}{1} by coronal
      X-rays shining down on the lower chromosphere at T$\sim$10$^4$~K
      (Zirin 1975; Laming \& Feldman 1993; Wahlstr$\o$m \& Carlsson 1994).
\item The \ion{O}{5}~$\lambda$1371 line suggests a very small emission
      measure. The line emissivity in Landini \& Monsignori Fossi (1990)
      suggests that the line should be an order of magnitude {\em brighter}
      than is observed. The \ion{O}{5} emission measure is inconsistent
      with the \ion{O}{4} and \ion{O}{6} line emission measures.
\end{itemize}

Note that the strong transition region resonance lines of \ion{Si}{4},
\ion{C}{4}, and \ion{N}{5} suggest emission measures somewhat above the
minimum values indicated by other ions.
We suggest that the excess flux, that above the minimum in the
emission measure diagram, may arise in the spatially-extended gas near the 
co-rotation radius.

\section{The Interstellar Absorption Features\label{sec-mgism}}

There is considerable evidence that the Sun lies inside a small cloud,
called the Local Interstellar Cloud (LIC), of 
warm partially ionized gas moving with a single--valued bulk flow velocity 
of --26 km s$^{-1}$ from Galactic coordinates $l^{\rm II} = 186^{\circ}$ 
and $b^{\rm II} = -16^{\circ}$, roughly 
the direction of the Sco--Cen association (Crutcher 1982;
Lallement et al.\/ 1995; Redfield \& Linsky 2000).  
In spectra of 
nearby stars, narrow interstellar absorption features with velocities equal 
to the projected velocity
of the LIC vector are identified as absorption from this cloud along the 
line of sight to the star.  Other nearby clouds that have been identified
include the G~Cloud, so named because it lies in the direction of the Galactic
center (Lallement \& Bertin 1992), and clouds in the direction of the
North Galactic Pole and South Galactic Pole (Redfield \& Linsky 2000).
Observations of
$\alpha$~Cen and 36~Oph indicate that the interface between the LIC and the
G~Cloud is located $<$~0.19~pc from the Sun roughly in 
the direction of the Galactic Center (Linsky \& Wood 1996; Wood, Linsky,
\& Zank 2000).
A three--dimensional
model of the LIC was developed by Redfield \& Linsky (2000)
based on 32 lines of sight
using HST, EUVE, and ground based \ion{Ca}{2} spectra.  In their model, the LIC
extends about 5 pc in the anti--Galactic Center direction,
and a fraction of a parsec in the Galactic Center direction, so that the Sun
is located just inside the very edge of the LIC.  For the line of sight
toward AB~Dor, the model predicts the presence of very little LIC material.

\begin{figure}
\plotone{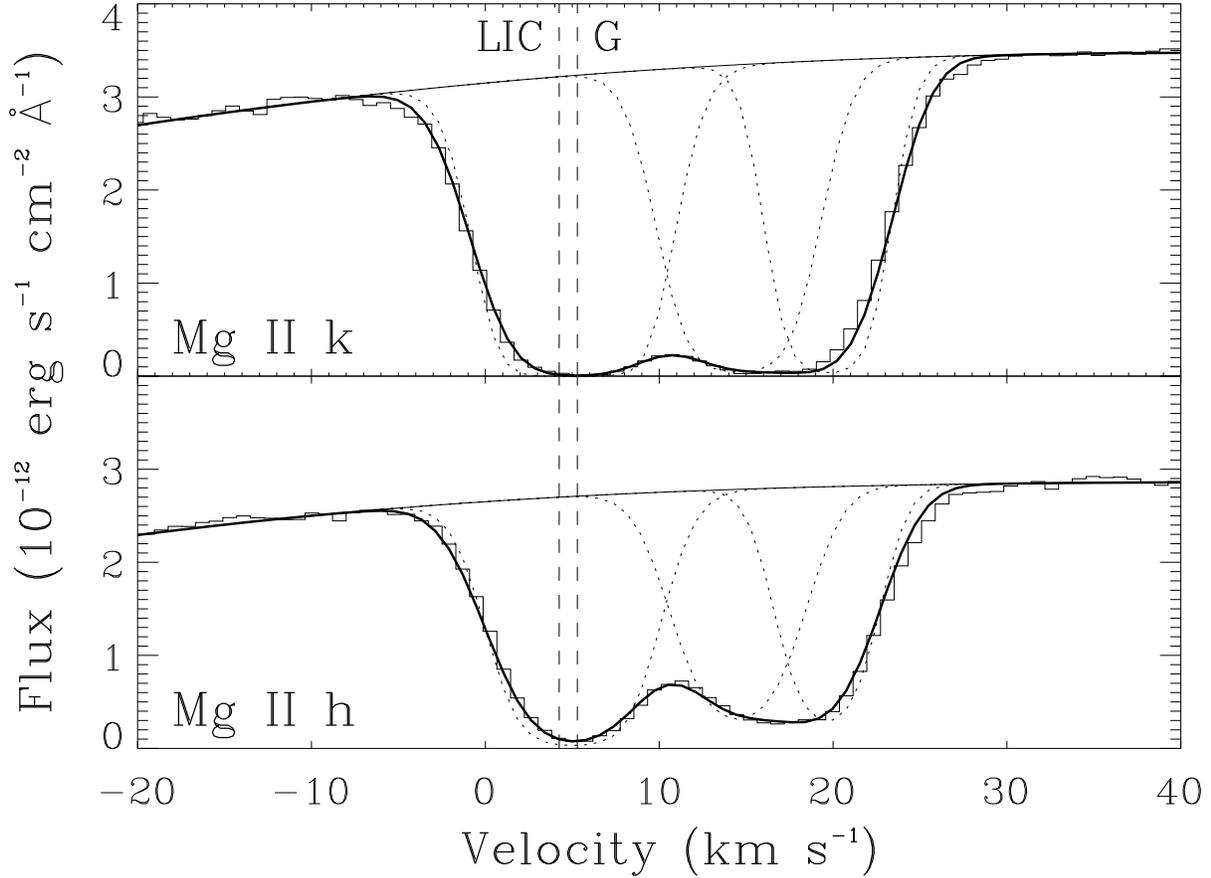}
\caption{The ISM absorption features observed in the 
\protect\ion{Mg}{2} lines. 
The velocity scale is heliocentric.
The data
are shown in histogram form.  Our best fit to the ISM lines is also shown
(see Table~\ref{mtbl} for the fit parameters).  
Fifth--degree polynomial fits to a mirrored line
profile ($thin$ $solid$ $lines$) have been used to interpolate the intrinsic
\protect\ion{Mg}{2} line profiles in the spectral regions where
the ISM absorption is located. 
The absorption profiles of the three interstellar absorption components are
shown as dotted lines, 
and the convolution of the profiles with the instrumental profile is shown
as a thick solid line.  Dashed vertical lines show the projected velocities
of the
LIC and G~cloud flow vectors for the AB~Dor line of sight.}\label{mgismfig}

\end{figure}

The \ion{Mg}{2} lines of AB~Dor contain
two broad interstellar absorption features. 
We fit the ISM features in the
mean \ion{Mg}{2}~$h$ and \ion{Mg}{2}~$k$ lines
simultaneously to maximally constrain the heliocentric velocity $v$ (km
s$^{-1}$), the Doppler parameter $b$ (km s$^{-1}$), and the \ion{Mg}{2} column
density 
$\log N_{\rm Mg II}$ 
($\log \rm cm^{-2}$).
The best fit is illustrated
in Figure~\ref{mgismfig}.  To estimate the ``continuum'' against which the
absorption features are measured, we use observed fluxes
on either side of the
absorption features and fluxes derived from mirroring the red--side of the line
together to fit a fifth--degree polynomial.  This ``continuum'' is shown in
Figure~\ref{mgismfig} as the thin solid line.  For a given set of interstellar
parameters ($v$, $b$, $N_{\rm{Mg II}}$),
the profiles for the interstellar
absorption features were computed and then convolved with the echelle--B LSA
instrumental profile (Gilliland \& Hulbert 1993).
In Figure~\ref{mgismfig} the absorption
profiles of the individual ISM components are shown as dotted lines,
and the convolution of these components
with the instrumental profile is shown as the thick solid line.  A scattered
light correction was made by simply subtracting a small flux level across the
whole spectrum.  Numerous fits were made with different estimates of the
scattered light
correction factors in order to estimate the systematic errors.  For the most
part, the uncertainties are dominated by the random error.  The derived
parameters for these interstellar absorption components 
are listed in Table~\ref{mtbl}. 

\placetable{mtbl}
\placefigure{mgismfig}

The two broad interstellar absorption features in the \ion{Mg}{2} lines can be
resolved into three different components -- one with a heliocentric velocity of
5.17~km~s$^{-1}$ and the other two with velocities of 14.5 and
19.6~km~s$^{-1}$.  A two component model inaccurately fits the high velocity
feature in the \ion{Mg}{2}~$k$ line,
so a third component was added to adequately fit
the flattened appearance of the high velocity absorption feature between 14 and
19~km~s$^{-1}$.  Since the quality--of--fit does not improve
significantly with the addition of a
fourth component, the data do not require more than
three components.  With a Galactic longitude of $l^{\rm II} = 275^{\circ}$ and
a latitude of $b^{\rm II} = -33^{\circ}$, the projected LIC velocity toward
AB~Dor should be 4.25~km~s$^{-1}$ and the corresponding G~Cloud velocity
should be 5.29~km~s$^{-1}$.  Because the projected velocities of the LIC and
G~Cloud are so similar in value, it is difficult to interpret the observed
5.17~km~s $^{-1}$ component as being associated with only one or the other
cloud. 
The fit itself seems to favor the G~Cloud projected velocity vector, and based
on the LIC model and observations of other nearby stars ($\alpha$~Cen and
36~Oph) we expect very little LIC material along this line of sight.  The LIC
model predicts a hydrogen column density of $\log N_{\rm HI}$~=~16.88
(Redfield \& Linsky 2000).  Using a typical magnesium  depletion for the LIC of
D(Mg)~$\sim$~--1.1~$\pm$~0.2 from Piskunov et al. (1997),
the predicted \ion{Mg}{2} column
density contribution of the LIC in the direction of AB~Dor is only
$\log N_{\rm Mg II}$~$\sim 11.4\ \pm\ 0.2$,
where $\rm{D(Mg)} = \log(\it N_{\rm{Mg II}}/\it
N_{\rm{HI}}) - \log(\rm{Mg/H})_{\odot}$.  Therefore, the LIC should be a small
contributor to the observed 5.17~km~s$^{-1}$ interstellar absorption
component.  However, if this absorption component is associated with the
G~Cloud, the large Doppler parameter, $b=3.8$~km~s$^{-1}$, implies a large
turbulent velocity.  If we assume a temperature typical for the G~Cloud,
$T=5650~\rm K$, based on observations of $\alpha$~Cen and 36~Oph
(Linsky \& Wood 1996; Wood, Linsky, \& Zank 2000),
the turbulent velocity is then $\xi=3.3~\rm{km~s^{-1}}$,
where we have used the equation $b^2 = (0.0165T/A + \xi^2)$ with  $A = 24.3$
for magnesium.  Other turbulent velocities measured for the G~Cloud by
Linsky \& Wood (1996) and Wood, Linsky, \& Zank (2000)
have significantly lower magnitudes.  This high
turbulent velocity could be caused by a shearing of the gas, possibly near the
interface of the  cloud.  However, it is more likely that because the LIC and
G~Cloud projected velocities are so close, we are unable to differentiate
between them in the spectrum.  Both clouds may be contributing to the broad
absorption feature, leading to the appearance of a large Doppler parameter and
turbulent velocity if the absorption is ascribed to only one cloud.  The
observed 14.5 and 19.6~km~s$^{-1}$ components have not yet been identified with
any previously detected cloud. 

\section{Summary}


We have presented the overall quiescent characteristics of AB Doradus
as it appeared during our 1994 November HST/GHRS observations.
The transition region and \ion{Mg}{2} line profiles provide
evidence for significant amounts of material co-rotating with the star out
to at least 3 stellar radii with temperatures extending
from the lower chromosphere to at least 10$^5$K.
This picture supports a model with extended gas confined near the
Keplerian co-rotation radius (2.8 stellar radii) by large-scale magnetic fields.
We surmise that most of the detected emission, the
$\sim$60$-$80\% emitted in the narrow line component, arises in high
density, small scale height regions close to the photosphere.
The extended material may well
be at a lower density, as indicated by the low density derived from
the $\frac{\rm{Si III]}~\lambda 1892}{\rm{C III]}~\lambda 1908}$ line
ratio, since both lines are depleted at high density (with the \ion{C}{3}]
line depleted faster). Formation of the \ion{C}{3}]~$\lambda$~1908 line
in the low density extended gas would explain both the broad line profile
and the large flux.

These quiescent spectra alone contain
a wealth of information, and show that
high signal-to-noise, high dispersion spectra of magnetically
active stellar chromospheres and coronae provide important insights into
the spatial morphology of the stellar magnetospheres.
In particular, we can determine the emission measure distribution, 
the atmospheric densities, 
the spatial extent of the atmosphere, 
and the properties of the interstellar medium along the line of sight.
High signal-to-noise, high dispersion UV spectra of active stars
provide our best indication of the conditions from which our present day
solar atmosphere evolved.

\acknowledgements
This paper is based on observations made with the NASA/ESA
{\it Hubble Space Telescope}, obtained at the Space Telescope Science Institute,
which is operated by the Association of Universities for Research in
Astronomy, Inc., under NASA contract NAS 5-26555. This research was supported
primarily by NASA grant NAG51862 to SUNY Stony Brook. 
We also acknowledge support to the University of Colorado through grant 
S-56460-D.
We thank the many people
who contributed to the success of the HST and GHRS programs.

\clearpage

\clearpage

\clearpage

\clearpage

\clearpage

\clearpage

\clearpage

\clearpage

\clearpage

\clearpage

\clearpage
\begin{deluxetable}{lrrlrrr}
\tablecolumns{7}
\tablewidth{380pt}
\protect\tablecaption{GHRS Observation Log\label{tbl-obslog}}
\tablehead{
\colhead{Root} & \colhead{Date} & \colhead{UT} & \colhead{Grating} 
               & \colhead{time} & \colhead{Phase\tablenotemark{a}}
               & \colhead{Notes\tablenotemark{b}}\\
 & (1994 Nov) & \colhead{(start)} & & \colhead{(sec)}}
\startdata
Z2I60107 & 14 & 23: 4:20  & Ech-B &  256 & 0.930 & 1\\
Z2I60109 & 14 & 23:12:23  & G160M & 1228 & 0.941 & 2\\
Z2I6010B & 14 & 23:36: 4  & Ech-B &  256 & 0.973 & 1\\
Z2I6010D & 14 & 23:43:51  & G160M & 1228 & 0.983 & 3\\
Z2I6010F & 15 &  0: 8:38  & Ech-B &  256 & 1.017 & 1\\
Z2I6010H & \nodata   & \nodata   & G160M & 1228 & \nodata & 7\\ 
Z2I6010M & 15 &  0:59: 2  & Ech-B &  256 & 1.085 & 1\\ 
Z2I6010O & 15 &  1: 7: 5  & G160M & 1228 & 1.096 & 2\\
Z2I6010Q & 15 &  1:30:46  & Ech-B &  256 & 1.128 & 1\\
Z2I6010S & 15 &  1:38:33  & G160M & 1228 & 1.138 & 3\\
Z2I6010U & 15 &  2: 5: 2  & Ech-B &  256 & 1.174 & 1\\
Z2I6010W & \nodata   & \nodata   & G160M & 1228 & \nodata & 7\\ 
Z2I60111 & 15 &  2:53:50  & Ech-B &  256 & 1.240 & 1\\
Z2I60113 & 15 &  3: 1:47  & G160M & 1228 & 1.250 & 2 \\
Z2I60115 & 15 &  3:25:34  & Ech-B &  256 & 1.282 & 1\\
Z2I60117 & 15 &  3:33:21  & G160M & 1228 & 1.293 & 3\\
Z2I60119 & 15 &  3:57:16  & Ech-B &  256 & 1.325 & 1\\
Z2I6011B & 15 &  4: 5:17  & G160M & 1228 & 1.336 & 4 \\
Z2I6011G & 15 &  4:48:38  & Ech-B &  256 & 1.394 & 1\\
Z2I6011I & 15 &  4:56:35  & G160M & 1228 & 1.405 & 2\\
Z2I6011K & 15 &  5:20: 8  & Ech-B &  256 & 1.437 & 1\\
Z2I6011M & 15 &  5:28:17  & G160M & 1228 & 1.448 & 3\\
Z2I6011O & 15 &  5:52: 4  & Ech-B &  256 & 1.480 & 1\\
Z2I6011Q & 15 &  5:59:58  & G160M & 1228 & 1.491 & 5\\
Z2I6011V & 15 &  6:43:20  & Ech-B &  256 & 1.549 & 1\\
Z2I6011X & 15 &  6:51: 9  & G160M & 1228 & 1.560 & 2\\
Z2I6011Z & 15 &  7:14:50  & Ech-B &  256 & 1.592 & 1\\ 
Z2I60121 & 15 &  7:23:11  & G160M & 1228 & 1.603 & 3\\
Z2I60123 & 15 &  7:46:52  & Ech-B &  256 & 1.635 & 1\\ 
Z2I60125 & 15 &  7:54:47  & G160M & 1228 & 1.646 & 6\\
Z2I6012A & 15 &  8:38: 8  & Ech-B &  256 & 1.704 & 1\\ 
Z2I6012C & 15 &  8:45:51  & G160M & 1228 & 1.714 & 2\\
Z2I6012E & \nodata   & \nodata   & Ech-B &  256 & \nodata & 7\\ 
Z2I6012G & 15 &  9:17:35  & G160M & 1228 & 1.757 & 3\\
Z2I6012I & 15 &  9:41:31  & Ech-B &  256 & 1.790 & 1\\ 
Z2I6012K & 15 &  9:49:21  & G160M & 1228 & 1.800 & 4 \\
Z2I6012P & 15 & 10:52:38  & Ech-B &  256 & 1.885 & 1\\ 
Z2I6012R & \nodata   & \nodata   & G160M & 1228 & \nodata & 7\\    
Z2I6012T & 15 & 11:22:52  & Ech-B &  256 & 1.926 & 1\\ 
Z2I6012V & 15 & 12:45:21  & G140L &  844 & 2.038 & 8\\
Z2I6012W & 15 & 13: 3:29  & G140L &  844 & 2.062 & 9\\
\enddata
\tablenotetext{a}{Phase at start of observation. Zero phase occurs at
                 HJD 2444296.575. The rotation is 10440 plus the phase.}
\tablenotetext{b}{(1) \protect\ion{Mg}{2} h\& K; 
                  (2) \protect\ion{Si}{4}; (3) \protect\ion{C}{4};
                  (4) \protect\ion{Si}{3}], \protect\ion{C}{3}];
                  (5) \protect\ion{N}{5}; (6) \protect\ion{C}{2}; 
                  (7) Observation failed;
                  (8) low dispersion 1305\AA; (9) low dispersion 1570\AA}
\end{deluxetable}

\clearpage
\begin{deluxetable}{rrl}
\tablecolumns{3}
\tablewidth{380pt}
\protect\tablecaption{Low Dispersion Line Atlas\label{tbl-lines}}
\tablehead{
\colhead{$\lambda$} & \colhead{flux} & \colhead{Line Identification} \\
                    & \colhead{(10$^{-14}$ erg cm$^{-2}$~s$^{-1}$)}}
\startdata
1175.86 & 44.0 $\pm$ 0.4  & \ion{C}{3} $\lambda$1175.711\\
1189.86 & 1.43 $\pm$ 0.28 & \ion{S}{3} $\lambda$1190.17; 
                            \ion{Si}{2} $\lambda$1190.42\\
1194.41 & 1.68 $\pm$ 0.29 & \ion{Si}{2} $\lambda$1193.29; 
                            \ion{S}{3} $\lambda\lambda$1194.06,1194.46\\
1197.56 & 0.40 $\pm$ 0.18 & \ion{Si}{2} $\lambda$1197.39\\
1201.13 & 2.77 $\pm$ 1.14 & \ion{N}{1} $\lambda\lambda$1200.22,1200.71;
                            \ion{S}{3} $\lambda\lambda$1200.97,1201.73 \\
1206.68 & 24.9 $\pm$ 0.7  & \ion{Si}{3} $\lambda$1206.51 \\
1215.91 & \nodata         & \ion{H}{1} ($\earth$)\\
1238.99 & 13.7 $\pm$ 0.28 & \ion{N}{5} $\lambda$1238.82 \\
1243.05 & 6.62 $\pm$ 0.25 & \ion{N}{5} $\lambda$1242.80 \\ 
1247.81 & 0.28 $\pm$ 0.16 & \ion{C}{3} $\lambda$1247.38\\
1253.96 & 0.31 $\pm$ 0.10 & \ion{Si}{2} $\lambda$1253.80\\
1259.57 & 0.96 $\pm$ 0.09 & \ion{Si}{2} $\lambda$1259.53\\ 
1261.08 & 1.01 $\pm$ 0.08 & \ion{C}{1} $\lambda$1261.3\\  
1264.81 & 1.46 $\pm$ 0.09 & \ion{Si}{2} $\lambda$1264.74\\
1266.50 & 0.74 $\pm$ 0.09 & \ion{C}{1} $\lambda$1266.42\\
1275.04 & 0.52 $\pm$ 0.07 & \nodata\\  
1277.33 & 1.19 $\pm$ 0.08 & \ion{C}{1} $\lambda$1277.5 \\
1280.14 & 0.98 $\pm$ 0.09 & \ion{C}{1} $\lambda$1280.5 \\
1288.77 & 0.64 $\pm$ 0.15 & \ion{C}{1} $\lambda$1288.42 \\
1294.87 & 0.70 $\pm$ 0.20 & \ion{Si}{3} $\lambda$1294.54 \\
1296.77 & 0.46 $\pm$ 0.28 & \ion{Si}{3} $\lambda$1296.72 \\
1299.09 & 1.31 $\pm$ 0.30 & \ion{Si}{3} $\lambda$1298.89 \\
1301.16 & 0.64 $\pm$ 0.07 & \ion{Si}{3} $\lambda$1301.15 \\
1302.33 & 5.27 $\pm$ 0.10 & \ion{O}{1} $\lambda$1302.16 \\
1304.97 & 5.27 $\pm$ 0.15 & \ion{O}{1} $\lambda$1304.86 \\
1306.12 & 7.06 $\pm$ 0.15 & \ion{O}{1} $\lambda$1306.03 \\
1309.59 & 0.90 $\pm$ 0.08 & \ion{Si}{2} $\lambda$1309.28 \\
1311.53 & 0.70 $\pm$ 0.09 & \ion{C}{1} $\lambda$1311.36 \\
1313.35 & 0.30 $\pm$ 0.08 & \nodata \\ 
1316.67 & 0.19 $\pm$ 0.06 & \ion{S}{1} $\lambda$1316.54 \\
1319.46 & 0.13 $\pm$ 0.06 & \ion{N}{1} $\lambda$1319.00 \\
1323.99 & 0.57 $\pm$ 0.16 & \ion{S}{1} $\lambda$1323.52\\
1329.28 & 1.17 $\pm$ 0.07 & \ion{C}{1} $\lambda$1329.1\\
1334.74 & 22.8 $\pm$ 0.2  & \ion{C}{2} $\lambda$1334.53\\
1335.83 & 35.2 $\pm$ 0.2  & \ion{C}{2} $\lambda$1335.71\\
1351.76 & 1.34 $\pm$ 0.08 & \ion{Cl}{1} $\lambda$1351.66 \\
1354.16 & 1.90 $\pm$ 0.08 & \ion{Fe}{21} $\lambda$1354.08 \\
1355.71 & 1.83 $\pm$ 0.10 & \ion{O}{1} $\lambda$1355.60 \\
1358.91 & 0.92 $\pm$ 0.10 & \ion{C}{1} $\lambda$1357.13; 
                            \ion{O}{1} $\lambda$1358.51\\
1361.63 & 0.27 $\pm$ 0.05 & \nodata \\
1363.27 & 0.29 $\pm$ 0.06 & \nodata \\  
1364.44 & 0.37 $\pm$ 0.06 & \ion{C}{1} $\lambda$1364.16 \\
1370.27 & 0.14 $\pm$ 0.05 & \ion{N}{3} $\lambda$1369.99 \\ 
1371.53 & 1.06 $\pm$ 0.12 & \ion{O}{5}  $\lambda$1371.292\\
1393.89 & 21.3 $\pm$ 0.05 & \ion{Si}{4} $\lambda$1393.76 \\
1399.72 & 0.65 $\pm$ 0.10 & \ion{O}{4}] $\lambda$1399.77 \\
1401.34 & 1.24 $\pm$ 0.15 & \ion{O}{4}] $\lambda$1401.16 \\
1402.90 & 12.0 $\pm$ 0.13 & \ion{Si}{4} $\lambda$1402.77 \\
1407.68 & 0.24 $\pm$ 0.10 & \ion{O}{4}] $\lambda$1407.39 \\
1412.59 & 0.34 $\pm$ 0.10 & \ion{Fe}{2} $\lambda$1412.83 \\
1416.71 & 0.25 $\pm$ 0.08 & \ion{S}{4}] $\lambda$1416.93 \\
1424.04\tablenotemark{a} & 0.14 $\pm$ 0.07 & \ion{S}{4}] $\lambda$1423.9 \\
1425.41 & 0.73 $\pm$ 0.11 & \ion{S}{1} $\lambda$1425.03 \\
1432.43 & 1.07 $\pm$ 0.10 & \ion{C}{1} $\lambda$1432.11 \\
1459.14 & 0.11 $\pm$ 0.10 & \ion{C}{1} $\lambda$1459.03 \\
1463.72 & 1.02 $\pm$ 0.13 & \ion{C}{1} $\lambda$1463.33; 
                            \ion{Fe}{10} $\lambda$1463.50 \\
1468.00 & 0.79 $\pm$ 0.16 & \ion{Fe}{9} $\lambda$1467.06; 
                            \ion{C}{1} $\lambda$1467.4 \\
1473.12 & 0.88 $\pm$ 0.12 & \ion{S}{1}  $\lambda$1472.97 \\
1474.36 & 1.22 $\pm$ 0.13 & \ion{S}{1}  $\lambda$1474.00\\
1481.93 & 1.12 $\pm$ 0.13 & \ion{S}{1}  $\lambda$1481.67 \\
1483.27 & 0.33 $\pm$ 0.08 & \ion{S}{1}  $\lambda$1483.23 \\
1486.89 & 0.90 $\pm$ 0.13 & \ion{N}{4} $\lambda$1486.50;
                            \ion{S}{1} $\lambda$1487.15 \\
1492.98 & 0.85 $\pm$ 0.13 & \nodata \\ 
1526.89 & 2.14 $\pm$ 0.13 & \ion{Si}{2} $\lambda$1526.71\\
1533.66 & 2.71 $\pm$ 0.61 & \ion{Si}{2} $\lambda$1533.43\\
1540.84 & 0.22 $\pm$ 0.10 & \ion{Fe}{2}  $\lambda$1541.03\\
1542.26 & 0.27 $\pm$ 0.09 & \ion{C}{1} $\lambda$1542.18\\
1548.40 & 80.9 $\pm$ 0.7  & \ion{C}{4} $\lambda$1548.20 \\
1550.94 & 42.7 $\pm$ 0.6  & \ion{C}{4} $\lambda$1550.77 \\
1561.09 & 9.31 $\pm$ 0.24 & \ion{C}{1} $\lambda$1561.00 \\
1640.52 & 84.7 $\pm$ 0.89 & \ion{He}{2} $\lambda$1640.48 \\
1649.98 & 0.54 $\pm$ 0.19 & \ion{Fe}{2} $\lambda\lambda$1649.42,1649.57 \\
1657.33 & 38.0 $\pm$ 0.71 & \ion{C}{1}
                            $\lambda\lambda$1556.28,1556.97,1557.38,1558.01\\
1666.70 & 1.52 $\pm$ 0.19 & \ion{O}{3}] $\lambda$1666.15 \\
1670.86 & 5.87 $\pm$ 0.25 & \ion{Fe}{2} $\lambda$1670.74\\
1686.79 & 0.42 $\pm$ 0.19 & \ion{Fe}{2} $\lambda$1686.46\\
1697.03 & 2.40 $\pm$ 0.34 & \ion{Fe}{2} $\lambda$1696.79\\
1713.18 & 1.64 $\pm$ 0.13 & \ion{Fe}{2} $\lambda$1713.0\\
\enddata
\tablenotetext{a}{Wavelength fixed; blended with S I $\lambda$1425.03.}
\end{deluxetable}  

\clearpage
\begin{deluxetable}{llll}
\tablecolumns{4}
\tablewidth{380pt}
\protect\tablecaption{Lines in $\lambda$1900\AA\ Spectral Region\label{tbl-1900}}
\tablehead{
\colhead{$\lambda$\tablenotemark{a}} & \colhead{flux} & \colhead{FWHM} & 
                    \colhead{Line Identification}    \\
        & \colhead{(10$^{-14}$ erg cm$^{-2}$s$^{-1}$)} & \colhead{\AA}}
\startdata
1892.30 $\pm$ 0.01 & 1.51 $\pm$ 0.01 & 1.28 $\pm$ 0.05 &
        \protect\ion{Si}{3}] $\lambda$1892.030 \\
1893.58 $\pm$ 0.02 & 0.09 $\pm$ 0.02 & 0.26 $\pm$ 0.07 &
        \protect\ion{S}{1} $\lambda$1893.252 ? \\
1895.39 $\pm$ 0.02 & 0.52 $\pm$ 0.03 & 0.63 $\pm$ 0.09 &
        \protect\ion{S}{1} $\lambda$1895.459 \\
1900.48 $\pm$ 0.02 & 0.34 $\pm$ 0.01 & 0.81 $\pm$ 0.13 &
        \protect\ion{S}{1} $\lambda$1900.270\\
1901.74 $\pm$ 0.01 & 0.54 $\pm$ 0.01 & 0.94 $\pm$ 0.10 &
        \protect\ion{Si}{1} $\lambda$1901.338  \\ 
1907.34 $\pm$ 0.04 & 0.25 $\pm$ 0.02 & 0.88 $\pm$ 0.16 \\
1909.38 $\pm$ 0.04 & 1.1  $\pm$ 0.1  & 2.09 $\pm$ 0.15 &
        \protect\ion{C}{3}] $\lambda$ 1908.734\tablenotemark{b}\\
1915.05 $\pm$ 0.04 & 0.75 $\pm$ 0.03 & 0.87 $\pm$ 0.24 &
        \protect\ion{S}{1} $\lambda$1914.680 \\
\enddata
\tablenotetext{a}{Observed wavelength. The rest wavelengths are
                  0.19\AA\ smaller.}
\tablenotetext{b}{Broad line. See discussion in Section~\ref{sec-dens4}.}
\end{deluxetable}  

\clearpage
\begin{deluxetable}{lll}
\tablecolumns{3}
\tablewidth{380pt}
\protect\tablecaption{Weak Lines in the G160M $\lambda$1400\AA\
              Spectrum\label{tbl-1400}}
\tablehead{
\colhead{$\lambda$\tablenotemark{a}} & \colhead{flux} &
             \colhead{Line Identification} \\
                    & \colhead{(10$^{-14}$ erg cm$^{-2}$s$^{-1}$)}}
\startdata
1400.07 $\pm$ 0.04 & 0.5  $\pm$ 0.1 & \ion{O}{4}] $\lambda$1399.77 \\
1401.58 $\pm$ 0.03 & 1.2  $\pm$ 0.3 & \ion{O}{4}] $\lambda$1401.16 \\
1407.70 $\pm$ 0.02 & 0.26 $\pm$ 0.04 & \ion{O}{4}] $\lambda$1407.39 \\
\enddata
\tablenotetext{a}{Observed wavelength. The rest wavelengths are 0.14\AA\
smaller.}
\end{deluxetable}  

\clearpage
\begin{deluxetable}{llll}
\tablecolumns{4}
\tablewidth{380pt}
\protect\tablecaption{Density Diagnostics\label{tbl-dens}}
\tablehead{
\colhead{Lines} & \colhead{ratio} & \colhead{log n$_e$ (cm$^{-3}$)}
                & \colhead{log n$_e$ (cm$^{-3}$) \tablenotemark{a}} }
\startdata
$\frac{\rm{Si III}~\lambda 1206}{\rm{Si III]}~\lambda 1892}$ 
              & 16.5 $\pm$ 0.5 & 12.2 $\pm$ 0.1 \\
$\frac{\rm{Si III}~\lambda 1301}{\rm{Si III]}~\lambda 1892}$ 
              & 0.42 $\pm$ 0.05 & 12.5 $\pm$ 0.1 \\
$\frac{\rm{Si III}~\lambda\lambda 1294-1301}{\rm{Si III]}~\lambda 1892}$ 
              & 2.0 $\pm$ 0.2 & 12.2 $\pm$ 0.1 \\
$\frac{\rm{Si III}~\lambda\lambda 1294-1301}{\rm{Si III}~\lambda 1206}$ 
              & 0.12 $\pm$ 0.01 & 12.2~\ud{0.4}{0.8} \\
$\frac{\rm{C III}~\lambda 1247}{\rm{C III}]~\lambda 1908}$ & 0.25 $\pm$ 0.14 & 
              11.1 \ud{0.4}{0.2} \\
$\frac{\rm{C III}~\lambda 1175}{\rm{C III}~\lambda 1247}$ & 157 $\pm$ 89 & 
              9.1 \ud{1.1}{0.8} & 11.8 \ud{2.8}{0.8} \\
$\frac{\rm{C III}~\lambda 1175}{\rm{C III}]~\lambda 1908}$ & 39 $\pm$ 1 & 
              10.8 $\pm$ 0.1 & 11.1 $\pm$ 0.1 \\
$\frac{\rm{Si IV}~\lambda 1402}{\rm{C III}]~\lambda 1908}$
              & 10.6 $\pm$ 0.5 & 11.5 $\pm$ 0.1 \\
$\frac{\rm{Si III]}~\lambda 1892}{\rm{C III}]~\lambda 1908}$ 
              & 1.3 $\pm$ 0.1 & 10.5 $\pm$ 0.1 \\
\enddata
\tablenotetext{a}{Density computed after doubling the
                  \protect\ion{C}{3}~$\lambda$1175 flux.
See Section~\ref{sec-dens2}.}
\end{deluxetable}  

\clearpage
\begin{deluxetable}{lccc}
\tablecolumns{4}
\tablewidth{380pt}
\protect\tablecaption{Fit Parameters for the Mg II Interstellar Components\label{mtbl}}
\tablehead{ Component & \colhead{$v$} & \colhead{$b$} & 
   \colhead{$\log N_{\rm MgII}$} \\
 & \colhead{(km s$^{-1}$)}&\colhead{(km s$^{-1}$)}&\colhead{$\log 
   (\rm cm^{-2})$}}
\startdata
1  & \phn5.2 $\pm$ 0.1 & 3.8 $\pm$ 0.1 & 13.13 $\pm$ 0.06 \\
2  &    14.5 $\pm$ 0.8 & 3.0 $\pm$ 0.8 & 12.8\phn $\pm$ 0.4\phn \\
3  &    19.6 $\pm$ 0.7 & 2.8 $\pm$ 0.5 & 12.7\phn $\pm$ 0.2\phn
\enddata
\end{deluxetable}  

\end{document}